\documentclass{myajour}
\usepackage{amssymb,jfa}

\begin{document}

%
\journame{Journal of Functional Analysis}
%
%
\authorrunninghead{Broderix, Leschke and M{\"u}ller}
\titlerunninghead{Integral kernels for unbounded Schr{\"o}dinger semigroups}
%
%


\title{Continuous integral kernels for unbounded Schr{\"o}dinger semigroups and
  their spectral projections} 

\author{Kurt Broderix\thanks{Deceased (12 May 2000).}}
\affil{Institut f{\"u}r Theoretische Physik, Georg-August-Universit{\"a}t,
  Friedrich-Hund-Platz 1,\\ D--37077 G{\"o}ttingen, Germany}

\author{Hajo Leschke}
\affil{Institut f{\"u}r Theoretische Physik, Universit{\"a}t
  Erlangen-N{\"u}rnberg, Staudtstra{\ss}{}e 7,\\ D--91058 Erlangen, Germany}
\email{hajo.leschke@physik.uni-erlangen.de}

\and

\author{Peter M{\"u}ller\thanks{Partially supported by SFB 602 of the Deutsche 
    Forschungsgemeinschaft.}}
\affil{Institut f{\"u}r Theoretische Physik, Georg-August-Universit{\"a}t,
  Friedrich-Hund-Platz 1,\\ D--37077 G{\"o}ttingen, Germany}
\email{peter.mueller@physik.uni-goettingen.de}

\newcommand{\sth}{$\mkern1mu ^{\mathrm{th}}$}

\dedication{Dedicated to Volker Enss on the occasion of his
  60\sth{} birthday}

\abstract{By suitably extending a Feynman-Kac formula of Simon 
  [Canadian Math. Soc. Conf. Proc. \textbf{28} (2000), 317--321], we
  study one-parameter semigroups  
  generated by (the negative of) rather general Schr{\"o}dinger
  operators, which 
  may be unbounded from below and include a magnetic vector potential.
  In particular, a common domain of essential self-adjointness for such a 
  semigroup is specified. Moreover, each member of the
  semigroup is proven to be a maximal Carleman operator with a continuous
  integral kernel given by a Brownian-bridge expectation. 
  The results are used to show that
  the spectral projections of the generating Schr\"odinger operator
  also act as Carleman operators with continuous integral
  kernels. Applications to Schr{\"o}dinger operators with rather
  general random scalar potentials include a
  rigorous justification of an integral-kernel representation of their
  integrated density of states -- a relation frequently used in the
  physics literature on disordered solids.}

\begin{article}

\tableofcontents
\bigskip

%

\zerosection{Introduction}

In non-relativistic quantum physics \cite{GaPa90, GaPa91, Thi02} a spinless
(charged) particle with 
$d$-dimensional Euclidean configuration space $\mathbb{R}^{d}$, which is
subjected to a scalar potential $V$, as well as to a magnetic field
derived from a vector potential $A$, is characterized by a
Schr\"odinger operator $H \equiv H(A,V)$. The latter is a
linear, self-adjoint, second-order partial-differential operator
acting on a dense domain in the Hilbert space
$\mathrm{L}^{2}(\mathbb{R}^{d})$ of Lebesgue square-integrable
functions $\psi$ on $\mathbb{R}^{d}$ \cite{CFKS87, BlEx94}. The
spectrum of $H$ corresponds physically to the possible values $E \in
\mathbb{R}$ of the particle's energy.  Useful information on a given
Schr\"odinger operator $H$ can be obtained by studying its semigroup
$\{\e^{-tH}\}_{t \ge 0}$. As was convincingly demonstrated by Carmona
\cite{Car79} and Simon \cite{Sim79a,Sim82}, this, in turn, can be done
very efficiently by using the Feynman-Kac(-It\^o) formula
\cite{Sim79a, ChZh95, Szn98, BrHuLe00}, which provides a probabilistic
representation of $\e^{-tH}\psi$ in terms of a Brownian-motion
expectation. Until present, the most systematic study along these lines is
that of Simon \cite{Sim82}. It covers mostly situations without a
magnetic field and where the scalar potential $V$ is assumed to be
Kato decomposable. The latter assumption assures in particular that
the operator $H$ is bounded from below and, hence, that
$\{\e^{-tH}\}_{t \ge 0}$ is a family of bounded operators.  Part of
the regularity results in \cite{Sim82} were recently generalized to
allow for rather general magnetic fields and an arbitrary open subset
of $\mathbb{R}^{d}$ as the configuration space \cite{BrHuLe00}.  For
additional regularity results see \cite{GuKo96}.

Some physically interesting situations, however, are modelled by scalar
potentials which are not Kato decomposable and lead to Schr\"odinger operators
that are unbounded from below. Here we only mention the Stark effect of atoms,
electronic properties of disordered solids and the physically different, but
mathematically closely related problem of classical diffusion in random media.
For the first situation one uses a scalar potential with a term linear in the
position \cite{AvHe77,CFKS87}, and for the latter two situations the 
realizations of a suitable random scalar potential \cite{Kir89, CaLa90,
  PaFi92, LeMuWa03, Szn98, GaKo00, GaKoMo00}.  Gaussian random potentials are
very popular examples thereof in the physics literature on disordered systems
\cite{ShEf84,Efe97, LeWo89}.  Since $H$ is unbounded from below in these
cases, the associated Schr\"odinger semigroup $\{\e^{-tH}\}_{t \ge 0}$
consists of unbounded operators. Among other things, the unboundedness of the
operator exponentials $\e^{-tH}$ brings up new kinds of questions concerning
domains, common cores for different $t$, etc. In fact, there are interesting
analytic results on semigroups of unbounded linear operators even on abstract
Hilbert and Banach spaces for more than two decades \cite{Nus70, Hug77, Fr80,
  KlLa81} (see also Thm.~4.9 in \cite{Da80}). However, it was only recently
that Simon \cite{Sim00} singled out a maximal class of negative scalar
potentials such that $H$ is unbounded from below, but given an arbitrarily
large (time) parameter $t>0$ the operator exponential $\e^{-tH}$ still acts as
an integral operator on functions $\psi$, which have sufficiently fast decay
at infinity, and $\e^{-tH}\psi$ is given by a Feynman-Kac formula.

The present paper is in the spirit of Simon's note \cite{Sim00}. 
By suitably extending his Feynman-Kac formula we aim to achieve a better understanding of 
rather general unbounded Schr\"odinger semigroups $\{\e^{-tH}\}_{t \ge 0}$ on 
$\mathrm{L}^2(\mathbb{R}^d)$, which have remained widely unexplored up to now. 
To this end we consider a large class of 
scalar potentials which allows for the same fall-off towards minus infinity 
at infinity as was considered in \cite{Sim00}. 
In addition, the presence of rather general magnetic fields is
admitted. Under these assumptions, we prove continuity of the 
Feynman-Kac-It\^{o} integral kernel $k_{t}$ of $\e^{-tH}$ and of the
image function $\e^{-tH}\psi$, provided that $t>0$ and $\psi$
has sufficiently fast decay at infinity. 
Moreover, we extend the Feynman-Kac-It\^o representation of
$\e^{-tH}\psi$ to all $\psi$ in the domain of the possibly unbounded operator
$\e^{-tH}$. This yields an alternative characterization of its domain
and renders 
$\e^{-tH}$ the maximal Carleman operator induced by the
integral kernel $k_{t}$. A theorem of
Nussbaum \cite{Nus70} is applied to identify a common operator
core for $\e^{-tH}$ for all $t \ge 0$. Lemma~\ref{bridgelemma} and
Theorem~\ref{fki} summarize these results.
Semigroup properties of the family $\{\e^{-tH}\}_{t \ge 0}$ 
are compiled in Theorem~\ref{fkigroup}.
Similar to Thm.~B.7.8 in \cite{Sim82}, we infer in
Theorem~\ref{intkernel} the existence and
continuity of integral kernels for certain bounded functions of $H$,
thereby allowing one to evaluate related traces in terms of
integral kernels. In particular, all this is true for any 
spectral-projection operator $\Chi_{I}(H)$ of $H$ associated with a Borel set
$I \subset \mathbb{R}$ which is bounded from above, see 
Corollary~\ref{speckernel}. Finally, the functional calculus is extended  
to integral kernels in Corollary~\ref{kernelcalculus}.
Applications to Schr{\"o}dinger operators with rather 
general random scalar potentials yield a rigorous justification
of some statements which are frequently used in the
physics literature on disordered systems.
Corollary~\ref{doscor} delivers an integral-kernel representation of the 
integrated density of states and Corollary~\ref{avsemi}, respectively
its particularization to 
Gaussian random scalar potentials in Corollary~\ref{gauss}, concerns
properties of the integral kernel of the averaged
semigroup. 

The paper is organized as follows. Section~\ref{ressection} contains
the basic notions, the precise formulations of the results mentioned
in the previous paragraph and various comments.   
Sections~\ref{bridgeproof} to~\ref{rpotproof} are devoted to the proofs.


%
\section{Results and Comments}
\label{ressection}

\subsection{Basic Notation and Definitions} 

As usual, let $\mathbb{N} := \{1,2,3,\ldots\}$ denote the set of natural
numbers. Let $\mathbb{R}$, respectively $\mathbb{C}$,
denote the algebraic field of real, respectively complex numbers 
and let $\mathbb{Z}^d$ be the simple cubic unit-lattice in $d$ dimensions, 
$d\in\mathbb{N}$.  We fix a Cartesian co-ordinate system in
$d$-dimensional Euclidean space $\mathbb{R}^{d}$ and define an open
cube in $\mathbb{R}^{d}$ as a translate of the
$d$-fold Cartesian product $I\times\dots\times I$ of
an open interval $I\subseteq\mathbb{R}$. In particular, 
$\Lambda_{\ell}(x)$ stands for the open cube in $\mathbb{R}^{d}$ with
edge length $\ell >0$ and centre 
$x=(x_{1},\ldots,x_{d})\in\mathbb{R}^{d}$. The Euclidean scalar
product $x\cdot y:= \sum_{j=1}^{d}x_{j}y_{j}$ of $x,y\in\mathbb{R}^{d}$
induces the Euclidean norm $|x|:=(x\cdot x)^{1/2}$. 

We denote the volume of a Borel subset $\Lambda\subseteq\mathbb{R}^{d}$
with respect to the $d$-dimensional Lebesgue measure as $|\Lambda| :=
\int_{\Lambda}\d{}x = \int_{\mathbb{R}^d}\!\d{}x \, \Chi_{\Lambda}(x)$,
where $\Chi_{\Lambda}$ stands for the indicator function of $\Lambda$.
In particular, if $\Lambda$ is the strictly positive half-line,
$\Theta := \Chi_{]\, 0 , \infty[}$ denotes the left-continuous Heaviside 
unit-step function.  

The Banach space $\mathrm{L}^{p}(\mathbb{R}^{d})$, $p \in [1,\infty]$, 
consists of all Borel-measurable complex-valued functions 
$f: \mathbb{R}^{d}\to\mathbb{C}$ which are identified if their values
differ only on a set of Lebesgue measure zero and which 
possess a finite norm $\|f\|_{p} := \bigl(\int_{\mathbb{R}^{d}}\! 
\d{}x\,|f(x)|^{p} \bigr)^{1/p} < \infty$, if $p < \infty$,  
and $ \| f\|_\infty := \esssup_{x\in\mathbb{R}^{d}} |f(x)| <
\infty$, if $p= \infty$. We recall that $\mathrm{L}^{2}(\mathbb{R}^{d})$ is a
separable Hilbert space with scalar product 
$\langle \cdot,\cdot \rangle$ given by $\langle f, g \rangle :=
\int_{\mathbb{R}^{d}}\d{}x\, f^{*}(x)\, g(x)$. Here the 
star denotes complex conjugation and the function $f^*$ is defined pointwise 
by $f^*(x):=(f(x))^*$. We write 
$f \in \mathrm{L}^{p}_\mathrm{loc}(\mathbb{R}^{d})$, if 
$f \Chi_{\Lambda} \in \mathrm{L}^{p}(\mathbb{R}^{d})$ for any bounded
Borel set $\Lambda\subset \mathbb{R}^{d}$. The uniform local Lebesgue spaces
$\mathrm{L}^{p}_{\mathrm{unif, loc}}(\mathbb{R}^{d})$ consist of all those
$f\in\mathrm{L}^{p}_\mathrm{loc}(\mathbb{R}^{d})$ for which 
$\sup_{x\in\mathbb{Z}^{d}} \| f\Chi_{\Lambda_{1}(x)}\|_{p} < \infty$.
The Kato class \cite{Kat72,AiSi82,Voi86,GuKo96} over $\mathbb{R}^{d}$ may be
defined as the vector space
$\mathcal{K}(\mathbb{R}^d ) := \big\{ f \in
\mathrm{L}^{1}_\mathrm{loc}(\mathbb{R}^{d}): \lim_{t\downarrow 0 }
\varkappa_t(f) = 0 \big\}$,  
where $\varkappa_t(f) := \sup_{x \in \mathbb{R}^d} \int_0^t \! \d s \, 
\int_{\mathbb{R}^d} \! \d\xi \, \e^{- | \xi |^2 }   | f(x +  \xi
\sqrt{s}) | $. It obeys the inclusion $\mathcal{K}(\mathbb{R}^d )
\subseteq \mathrm{L}^{1}_{\mathrm{unif, loc}}(\mathbb{R}^{d})$
with equality if $d=1$.
We say that $f$ belongs to $\mathcal{K}_{\mathrm{loc}}(\mathbb{R}^{d})$, if
$f\Chi_{\Lambda}\in\mathcal{K}(\mathbb{R}^d )$ for any bounded Borel set 
$\Lambda\subset \mathbb{R}^{d}$. Moreover, $f$ is called Kato decomposable,
in symbols $f\in\mathcal{K}_{\pm}(\mathbb{R}^{d})$, if
$\sup\{0,f\}\in\mathcal{K}_{\mathrm{loc}}(\mathbb{R}^{d})$ and 
$\sup\{0,-f\}\in\mathcal{K}(\mathbb{R}^{d})$.
Finally, $\mathcal{C}^{\infty}_{0}(\mathbb{R}^{d})$ is the 
vector space of all functions $f: \mathbb{R}^{d}\to\mathbb{C} $ which are
arbitrarily often differentiable and have compact supports $\mathrm{supp} f$. 

The absolute value of a closed operator 
$F: \dom(F) \to \mathrm{L}^2(\mathbb{R}^{d})$, with dense
domain of definition $\dom(F) \subseteq \mathrm{L}^2(\mathbb{R}^{d})$ 
and Hilbert adjoint $ F^{*}$, is the positive operator 
$|F| := (F^{*}F)^{1/2}$. The (uniform) norm of a bounded operator $F :
\mathrm{L}^2(\mathbb{R}^{d}) \to \mathrm{L}^2(\mathbb{R}^{d})$ is  
defined as $\| F \| := \sup\big\{ \| F f \|_2 \, : \, 
  f \in \mathrm{L}^2(\mathbb{R}^{d})\, ,  \| f \|_2 =1 \big\} $.

\begin{definition}
  \label{basicdef}
  Let $d\in\mathbb{N}$. A \emph{vector potential} $A$ is a Borel-measurable,
  $\mathbb{R}^{d}$-valued function on $\mathbb{R}^{d}$ and a \emph{scalar
  potential} $V$ is a Borel-measurable, $\mathbb{R}$-valued function on
  $\mathbb{R}^{d}$. Furthermore,
  \begin{indentnummer}
  \item[\ass{A}]
    a vector potential $A$ is said to satisfy property \ass{A}, if
    both its square $|A|^{2}$ and its divergence $\nabla\cdot A$ lie in
    the intersection
    $\mathrm{L}^{2}_{\mathrm{loc}}(\mathbb{R}^{d}) \cap
    \mathcal{K}_{\mathrm{loc}}(\mathbb{R}^{d})$. Here, $\nabla =
    (\partial_{1},\ldots,\partial_{d})$ stands for the gradient, which
    is supposed to act in the sense of distributions on
    $\mathcal{C}^{\infty}_{0}(\mathbb{R}^{d})$.  
  \item[\ass{C}] 
    a vector potential $A$ is said to satisfy property \ass{C}, if 
    there exist real constants $B_{jk} = -B_{kj}$, where $j,k\in
    \{1,\ldots,d\}$, such that
    \begin{equation}
      \label{constfield}
      A_{k}(x) = \frac{1}{2}\,\sum_{j=1}^{d} x_{j}\, B_{jk}
    \end{equation}
    for all $x\in\mathbb{R}^{d}$ and all $k\in \{1,\ldots,d\}$.
    In other words, $A$ generates a spatially constant magnetic field
    given by the skew-symmetric $d\times d$-matrix with entries
    $B_{jk}=\partial_{j}A_{k} - \partial_{k}A_{j}$.  
  \item[\ass{V}] 
    a scalar potential $V$ is said to satisfy property 
    \ass{V}, if it can be written as a sum
    \begin{equation}
      V = V_1 + V_2
    \end{equation}
    with $V_1$ being locally square-integrable and Kato decomposable,
    \begin{equation}
      V_1\in\mathrm{L}^{2}_{\mathrm{loc}}(\mathbb{R}^{d}) \cap
      \mathcal{K}_{\pm}(\mathbb{R}^{d}) \,,
    \end{equation}
    and $V_2$ obeying a sub-quadratic growth limitation in the following 
    sense: for every $\varepsilon >0$ there exists a finite constant 
    $v_\varepsilon >0$ such that 
    \begin{equation}
      \label{Wcond}
      |V_2(x)| \leq \varepsilon |x|^{2} + v_\varepsilon 
    \end{equation}
    for Lebesgue-almost all $x\in\mathbb{R}^{d}$.
  \end{indentnummer}
\end{definition}

\begin{remarks}
  \begin{nummer}
    \item 
      For one space dimension, $d=1$, there is no loss of
      generality in assuming $A=0$ on account of gauge equivalence.
    \item If $d \le 3$, then $\mathrm{L}^{2}_{\mathrm{loc}}
      (\mathbb{R}^{d})\subseteq\mathcal{K}_{\mathrm{loc}}(\mathbb{R}^{d})$.
    \item  
      Due to gauge equivalence we have contented ourselves in formulating the 
      constant-magnetic-field condition \ass{C} 
      in the Poincar\'e gauge \eqref{constfield}. 
    \item 
      Property \ass{C} implies property \ass{A}.
    \item \label{sim00ass}
      Property \ass{V} allows for a larger class of potentials than
      those considered in \cite{Sim00}. This is because \ass{V} requires 
      weaker local regularity properties. Yet, the crucial sub-quadratic
      growth limitation of $V(x)$ towards minus infinity as 
      $|x| \to \infty$ is identical. 
    \item Even though a quadratic growth limitation instead of the 
      stronger condition
      \eqref{Wcond} would still yield a self-adjoint Schr\"odinger semigroup,
      we do not consider such situations, because the corresponding 
      Feynman-Kac(-It\^o) formula would not hold for an arbitrarily large
      time parameter $t$, cf.\ Sect.~5.13  in \cite{ItMc74}.   
  \end{nummer}
\end{remarks}

We base the definition of Schr{\"o}dinger operators on the following
proposition, whose proof is an application of Thm.~2.5 in \cite{HiSt92}.

\begin{proposition}
  \label{Hesssa}
  Let $A$ be a vector potential with property \ass{A} and let $V$ be a scalar
  potential with property \ass{V}. Then the differential operator 
  \begin{equation}
    \label{Hess}
    \mathcal{C}_0^\infty(\mathbb{R}^d) \ni \varphi \mapsto 
    \frac{1}{2} \sum_{j = 1}^d ({\i} \partial_j + \hat{A}_j)^2 \, \varphi +
    \hat{V}\varphi 
  \end{equation}
  is essentially self-adjoint on $\mathrm{L}^2(\mathbb{R}^d)$. Here
  ${\i}=\sqrt{-1}$ denotes the imaginary unit and a superposed hat
  on a function indicates the corresponding multiplication operator.
\end{proposition}

\begin{definition}
  \label{Hdef}
  The self-adjoint closure of \eqref{Hess}  
  on $\mathrm{L}^2(\mathbb{R}^d)$ is called the \emph{(magnetic)
    Schr{\"o}\-dinger operator} and denoted by $H(A,V)$.   
\end{definition}

As suggested in \cite{Sim00}, we introduce vector spaces of $\mathrm{L}^{p}
(\mathbb{R}^{d})$-functions with a decay at infinity which is faster than 
that of some Gaussian function. These spaces are tailored
for the, in general, unbounded Schr\"odinger semigroup 
$\{\e^{-t H(A,V)}\}_{t\ge 0}$ with $V$ having property
\ass{V}. 

\begin{definition}
  For each $p \in[1,\infty]$ we set
  \begin{eqnarray}
    \mathrm{L}^{p}_{\mathrm{G}}(\mathbb{R}^{d}) := \bigg\{ \psi\in
    \mathrm{L}^{p}(\mathbb{R}^{d}) : & \mbox{there exists} &  \rho \in
    ]0,\infty[ \mbox{~~such that~~} \nonumber\\  
  && \hspace*{-3em} \int_{\mathbb{R}^{d}}\!\d x\; \e^{\rho |x|^{2}} \,
    |\psi(x)|^{p}  < \infty\bigg\}\,.
  \end{eqnarray}
\end{definition}

\begin{remarks}
  \begin{nummer}
  \item \label{lepsincl}
    H{\"o}lder's inequality yields the chain of inclusions
    \begin{equation}
      \mathrm{L}^{\infty}_{\mathrm{G}}(\mathbb{R}^{d}) \subseteq
      \mathrm{L}^{q}_{\mathrm{G}}(\mathbb{R}^{d}) \subseteq
      \mathrm{L}^{p}_{\mathrm{G}}(\mathbb{R}^{d}) \subseteq
      \mathrm{L}^{1}_{\mathrm{G}}(\mathbb{R}^{d}) \,,
    \end{equation}
    if $1\leq p \leq q \leq \infty$.
  \item \label{lepsdense}
    The space $\mathrm{L}^{p}_{\mathrm{G}}(\mathbb{R}^{d})$ is dense 
    in $\mathrm{L}^{p}(\mathbb{R}^{d})$ for any $p\in[1,\infty]$
    thanks to the inclusion 
    \begin{equation}
      \mathcal{C}_0^\infty(\mathbb{R}^{d}) \subset 
      \mathrm{L}^{p}_{\mathrm{G}}(\mathbb{R}^{d})\,.
    \end{equation}
  \end{nummer}
\end{remarks}

\subsection{Continuous integral kernels for unbounded Schr{\"o}dinger
  semigroups and their spectral projections}

As a preparation for the Feynman-Kac-It\^{o} formula \eqref{fkieq} in
Theorem~\ref{fki} below we need to recall the \emph{Brownian bridge}
in $\mathbb{R}^{d}$ associated with the starting point
$x\in\mathbb{R}^{d}$, the endpoint $y\in\mathbb{R}^{d}$ and the closed
time interval $[0,t]$, where $t>0$ is fixed but arbitrary. It may be
defined as the 
$\mathbb{R}^{d}$-valued 
stochastic process whose $d$ Cartesian components are independent and
have continuous realizations $[0,t] \ni s\mapsto b_{j}(s)
\in\mathbb{R}$, $j\in\{1,\ldots, d\}$. Moreover, the $j$-th component
$b_{j}$ is 
distributed according to the Gaussian probability measure
characterized by the mean function $[0,t]\ni s \mapsto x_{j} +
(y_{j}-x_{j})s/t$ and the covariance function $[0,t] \times [0,t] \ni
(s,s') \mapsto \min\{s,s'\} - ss'/t$, see e.g.\ \cite{Sim79a, Pro95, Szn98}.
We denote the joint (product) probability measure of
$b:=(b_{1},\ldots,b_{d})$ by 
$\mu_{x,y}^{0,t}$.
Given $t>0$, a vector potential $A$ with property \ass{A} and a scalar
potential $V$ with property \ass{V}, then
the \emph{Euclidean action functional}
\begin{equation}
  \label{action}
  S_{t}(A,V;b) := {\i}\int_{0}^{t}\!\d b(s)\cdot A(b(s)) +
  \frac{\i}{2}\,\int_{0}^{t}\!\d s\; (\nabla\cdot A)(b(s)) +
  \int_{0}^{t}\!\d s\; V(b(s))
\end{equation}
associated with these potentials 
is well defined for $\mu_{x,y}^{0,t}$-almost all paths $b$ of the
Brownian bridge. The first integral on the right-hand side of 
\eqref{action} is a stochastic line integral to be understood in the 
sense of It\^{o}. The other two integrals with random integrands are meant in
the sense of Lebesgue. The $\mu_{x,y}^{0,t}$-almost-sure existence of the 
integrals in \eqref{action} follows 
e.g.\ from Sects.~2 and~6 in \cite{BrHuLe00} and the estimate
\begin{equation}
  \quad\int\!\mu_{x,y}^{0,t}(\d b)\; \left| \int_{0}^{t}\!\d s\;
    V_{2}(b(s))\right| \le
  t v_\varepsilon + \varepsilon \int_{0}^{t}\!\d s 
  \int\!\mu_{x,y}^{0,t}(\d b)\; |b(s)|^2 < \infty \,.\quad
\end{equation}
The latter is valid for all $\varepsilon >0$ and relies on \eqref{Wcond},
Fubini's theorem and an explicit computation. As to the applicability of 
\eqref{Wcond} in this estimate, we have used the basic fact that 
for $\mu_{x,y}^{0,t}$-almost every path $b$ of the Brownian bridge the
set $\{ s \in [0,t] \, : \, b(s) \in \Lambda \}$ of time instances, for
which $b$ stays in a given Lebesgue-null set $\Lambda \subset
\mathbb{R}^d$, is itself of Lebesgue measure zero in $[0,t]$, that is,
$\int_{0}^{t} \d s\; \Chi_{\Lambda}\bigl(b(s)\bigr) =0$. 
We will make use of this fact in the following without further notice.

\begin{lemma}
  \label{bridgelemma}
  Let $A$ be a vector potential with property \ass{A} and let $V$ be a
  scalar potential with property \ass{V}. Finally, let $t>0$. Then
  \begin{nummer}
  \item \label{kerneldef}
    the function $k_{t} : \mathbb{R}^{d}\times\mathbb{R}^{d} 
    \to \mathbb{C}$, $(x,y)\mapsto k_{t}(x,y)$, where
    \begin{equation}
      \label{kern}
      k_{t}(x,y) := \frac{\e^{-|x-y|^{2}/(2t)}}{(2\pi t)^{d/2}}
      \int\!\mu_{x,y}^{0,t}(\d b)\;\e^{-S_{t}(A,V;b)} \,,
    \end{equation}
    is well defined in terms of a Brownian-bridge expectation,
    Hermitian in the sense that $k_{t}(x,y) = k_{t}^{*}(y,x)$ for all
    $x,y \in\mathbb{R}^{d}$,
    continuous and obeys the semigroup property  
    \begin{equation}
      \label{markov}
      k_{t+t'}(x,z)= \int_{\mathbb{R}^d}\!\d y\; k_t(x,y)\,k_{t'}(y,z)
    \end{equation}
    for all $x,z\in\mathbb{R}^{d}$ and all $t' >0$.
  \item  \label{kernelbound}
    for every $\delta >0$ there exists a finite constant $a_{t}^{(\delta)}>0$, 
    independent of $x,y\in\mathbb{R}^{d}$, such that the estimate 
    \begin{equation}
      \label{kbound}
      |k_{t}(x,y)| \leq a_{t}^{(\delta)} \exp\biggl\{-
      \frac{|x-y|^{2}}{4t} 
      + \delta |x|^{2} + \delta |y|^{2}\biggr\} 
    \end{equation}
    holds for all $x,y\in\mathbb{R}^{d}$. 
  \item \label{kCarleman}
    the function $k_t$ obeys
    \begin{equation}
      \label{kstark}
      k_t(x,\cdot) \in\mathrm{L}^\infty_{\mathrm{G}}(\mathbb{R}^d)
      \qquad \mbox{for all~~}
      x\in\mathbb{R}^d
    \end{equation}
    and thus has the Carleman property \eqref{carledef} below.
    Moreover, the mapping $\mathbb{R}^{d}\to\mathrm{L}^{2}(\mathbb{R}^{d})$, 
    $x \mapsto k_t(x,\cdot)$ is strongly continuous.
\end{nummer}
\end{lemma}

\begin{remarks}  
  \begin{nummer} 
  \item The lemma is proven in Section \ref{bridgeproof}. 
  \item Concerning the asserted continuity of $k_{t}$, the proof will
    even show that 
    the function $ ]0,\infty[ \times\mathbb{R}^{d}\times\mathbb{R}^{d} \ni
    (t,x,y) \mapsto k_{t}(x,y)$ is continuous. 
  \item The estimate (\ref{kbound}) corresponds to Thm.~2.1 in \cite{Sim00}.
  \item \label{Carleswap}
    Part~\itemref{kCarleman} of Lemma~\ref{bridgelemma} continues to hold 
    with $k_t(x,\cdot)$ replaced by $k_t(\cdot,x)$ thanks to the
    Hermiticity of $k_t$ (for \emph{all} $x,y\in\mathbb{R}^d$). 
  \item While \eqref{kstark} follows (directly) 
    from the estimate \eqref{kbound}, the weaker \emph{Carleman
      property} of $k_{t}$,
    \begin{equation}
      \label{carledef}
      k_t(x,\cdot) \in \mathrm{L}^2(\mathbb{R}^d) \mbox{~~~for 
        Lebesgue-almost all~~} x\in\mathbb{R}^d\,,
    \end{equation}
    is already a consequence
    of the semigroup property, the Hermiticity  and the continuity of $k_t$. 
  \end{nummer}
\end{remarks}

\begin{definition}
  \label{semidef}
  Let $H(A,V)$ be the Schr\"odinger operator of Definition~\ref{Hdef} and 
  let $t\in\mathbb{R}$. Then the operator exponential $\e^{-t H(A,V)}$ 
  is densely defined, self-adjoint and positive by the spectral
  theorem and the  
  functional calculus for unbounded functions of unbounded
  self-adjoint operators (see e.g.\ Chap.~5 in \cite{BlEx94}).
\end{definition}

We are now in a position to give a probabilistic representation of 
$\e^{-tH(A,V)}$ by a Feynman-Kac-It\^{o} formula.

\begin{theorem}
  \label{fki}
  Let $A$ be a vector potential with property \ass{A} and let $V$ be a scalar 
  potential with property \ass{V}. Moreover, let $t>0$ and let
  $\e^{-t H(A,V)}$ be given by Definition~\ref{semidef}. Then
  \begin{nummer}
  \item 
    \label{fkidomain}
    the domain of $\e^{-tH(A,V)}$ is given by
    \begin{equation}
      \dom\bigl(\e^{-tH(A,V)}\bigr)
      = \Bigl\{\psi\in\mathrm{L}^{2}(\mathbb{R}^{d}): 
      \int_{\mathbb{R}^{d}}\d{}y\; k_{t}(\cdot,y)\,\psi(y) \in
      \mathrm{L}^{2}(\mathbb{R}^{d}) \Bigr\} \qquad
      \label{domain}
    \end{equation}
    with $k_{t}$ defined in \eqref{kern}.
    Moreover, $\mathrm{L}^2_{\mathrm{G}}(\mathbb{R}^d) \subseteq 
    \dom\bigl(\e^{-tH(A,V)}\bigr)$ is an operator core for $\e^{-tH(A,V)}$. 
  \item 
    \label{fkiintegral}
    $\e^{-tH(A,V)}$ is the maximal Carleman operator induced by the 
    continuous integral kernel \eqref{kern}
    in the sense that
    \begin{equation}
      \label{fkieq}
      ~\qquad
      \e^{-tH(A,V)}\psi = \int_{\mathbb{R}^{d}}\!\d{}y\;
      k_{t}(\cdot,y)\, \psi(y)     
    \end{equation}
    for all $\psi\in \dom\bigl(\e^{-tH(A,V)}\bigr)$
    and that $k_t$ has the Carleman property \eqref{carledef}.
  \item 
    \label{fkicont}
    the image $\e^{-tH(A,V)}\psi$ of any
    $\psi\in\dom\bigl(\e^{-tH(A,V)}\bigr)$ has a continuous
    representative in $\mathrm{L}^{2}(\mathbb{R}^{d})$ given by the 
    right-hand side of \eqref{fkieq}. If even
    $\psi\in\mathrm{L}^2_{\mathrm{G}}(\mathbb{R}^d)$, then, in addition,
    $\e^{-tH(A,V)}\psi \in\mathrm{L}^{\infty}_{\mathrm{G}}(\mathbb{R}^d)$. 
  \end{nummer}  
\end{theorem}

\begin{remarks}
  \begin{nummer}
  \item The proof of Theorem~\ref{fki} is deferred to Section
    \ref{fkiproof}. 
  \item For the theory of Carleman operators we refer to 
    \cite{Sto70, AcGl81, Wei80}. We follow mostly the terminology and
    conventions of \cite{Wei80}.
  \item \label{lpwell}
    The right-hand side of \eqref{fkieq} maps even any $\psi\in 
    \mathrm{L}^{1}_{\mathrm{G}}(\mathbb{R}^{d})$
    (and hence any $\psi\in\mathrm{L}^{p}_{\mathrm{G}}(\mathbb{R}^{d})$ for
    all $p\in[1,\infty]$) to an element of 
    $\mathrm{L}^{\infty}_{\mathrm{G}}(\mathbb{R}^{d})$. This fact is
    well known for the free case $A=0$ and $V=0$. It extends to
    the general situation of Theorem~\ref{fki} simply by the basic
    estimate \eqref{kbound}.
  \item Theorem~\ref{fki} extends the main result of
    \cite{Sim00}, where the Feynman-Kac-It\^{o} formula \eqref{fkieq}
    was proven for $A=0$ and $\psi\in\mathrm{L}^2_{\mathrm{G}}
    (\mathbb{R}^{d})$ under somewhat more restrictive 
    assumptions on the scalar potential $V$, see Remark~\ref{sim00ass}.
  \item If $V_2=0$, then the scalar potential $V=V_{1}$ is Kato
    decomposable and $H(A,V_{1})$ therefore bounded from
    below. Regularity properties of 
    the associated \emph{bounded} Schr\"odinger semigroup
    $\{\e^{-tH(A,V_{1})}\}_{t\ge 0}$ are well known and have been 
    studied in great detail,
    see the seminal paper \cite{Sim82} and \cite{GuKo96} for the 
    non-magnetic case $A=0$. Part of these results were extended to
    situations with rather general vector potentials in \cite{BrHuLe00}. 
  \end{nummer}
\end{remarks}

So far we have been concerned with the (possibly unbounded) operator 
exponential $\e^{-tH(A,V)}$ for a fixed but arbitrary time parameter
$t\in ]0,\infty[$. Next
we compile some semigroup properties of the family 
$\{\e^{-tH(A,V)}\}_{t \ge 0}$. 

\begin{theorem}
  \label{fkigroup}
  Assume the situation of Theorem~\ref{fki}. Then the family
  $\{\e^{-tH(A,V)}\}_{t\ge 0}$ is a strongly continuous 
    (one-parameter) semigroup of self-adjoint
    operators generated by the Schr\"odinger
    operator $H(A,V)$ in the following sense:
  \begin{nummer}
  \item 
     the semigroup law 
    \begin{equation}
      \label{semigroupprop}
      \e^{-(t+t')H(A,V)}\psi= \e^{-tH(A,V)}\,
      \e^{-t'H(A,V)}\psi
    \end{equation}
    holds for all $t,t'\in[0,\infty[$ and all 
    $\psi\in\mathrm{L}^2_{\mathrm{G}}(\mathbb{R}^d)$.
  \item 
    the orbit mapping 
    $u_{\psi}: [0,\infty[ \rightarrow \mathrm{L}^{2}(\mathbb{R}^d)$,
    $t\mapsto u_{\psi}(t):= \e^{-tH(A,V)}\psi$ is strongly
    continuous (at $t=0$ only from the right) for all 
    $\psi\in\mathrm{L}^2_{\mathrm{G}}(\mathbb{R}^d)$. 
  \item 
    for every $\varphi\in\mathcal{C}_{0}^{\infty}(\mathbb{R}^{d})$ the
    orbit mapping $u_{\varphi}$ is strongly differentiable
    (at $t=0$ only from the right) and the unique solution of the
    linear initial-value problem 
    \begin{equation}
      \label{initval}
      \frac{\d}{\d t}\Phi(t) = -H(A,V) \Phi(t)\,, \qquad\quad
      \Phi(0) = \varphi\,,
    \end{equation}
    for a strongly differentiable (at $t=0$ only from the right)
    mapping $\Phi : [0, \infty[ \rightarrow \dom\bigl(H(A,V)\bigr)$,
    $t\mapsto \Phi(t)$.
  \end{nummer}
\end{theorem}

\begin{remarks}
  \begin{nummer}
  \item The proof of Theorem~\ref{fkigroup} is given in
    Section~\ref{fkiproof}.   
  \item Interesting analytic results on semigroups of unbounded operators on
    abstract Hilbert and Banach spaces were
    previously obtained in e.g.\ \cite{Nus70, Hug77, Fr80, KlLa81}. 
  \end{nummer}
\end{remarks}

In many situations it is useful to know that not only $\e^{-tH(A,V)}$
has a continuous integral kernel but also certain bounded functions 
of $H(A,V)$.

\begin{theorem} \label{intkernel}
  Assume the situation of Theorem~\ref{fki} and let 
  $F\in\mathrm{L}^{\infty}(\mathbb{R})$ be a bounded function with an
  at least exponentially fast decay at plus infinity in the sense that 
  the inequality 
  \begin{equation}
    \label{Fcond}
    |F(E)| \leq \gamma \min\bigl\{1, \e^{-\tau E}\bigr\}
  \end{equation}
  holds for Lebesgue-almost all $E\in\mathbb{R}$ with some 
  constants $\gamma,\tau\in]0,\infty[$. Furthermore, let
  $F\bigl(H(A,V)\bigr)$ be defined by the spectral theorem and the
  functional calculus. 
  Then
  \begin{nummer}
  \item
    \label{intkernelexist}
    $F\bigl(H(A,V)\bigr)$ is a bounded Carleman operator induced by
    the continuous integral kernel
    $f:\mathbb{R}^{d}\times\mathbb{R}^{d}\to\mathbb{C}$,  
    $(x,y)\mapsto f(x,y)$, where 
    \begin{equation}
      \label{fexeq}
      f(x,y) := \bigl\langle k_t(\cdot,x),\e^{2tH(A,V)}F\bigl(H(A,V)\bigr)
      k_t(\cdot,y)\bigr\rangle
    \end{equation}
    with arbitrary $t\in]0,\tau/2[$, in the sense that 
    \begin{equation}
      \label{fintkernel}
      F\bigl(H(A,V)\bigr)\psi = \int_{\mathbb{R}^{d}}\!\d{}y\;
      f(\cdot,y) \,\psi(y) 
    \end{equation}
    for all $\psi\in\mathrm{L}^{2}(\mathbb{R}^{d})$ and that $f$ has
    the Carleman property \eqref{carledef}.
  \item 
    \label{intkernelcont}
    the left-hand side of \eqref{fintkernel} has a continuous representative 
    in $\mathrm{L}^{2}(\mathbb{R}^{d})$, which is  given by the right-hand
    side of \eqref{fintkernel}.  
  \item 
    \label{intkernelHStrace}
    for every $w\in\mathrm{L}^{\infty}_{\mathrm{G}}(\mathbb{R}^{d})$
    the product
    $F\bigl(H(A,V)\bigr)\hat{w}$ is a Hilbert-Schmidt operator
    with squared norm given by
    \begin{equation}
      \label{HStrace}
      \mathrm{Trace}\bigl\{ \hat{w}^{*}\bigl|F\bigl(H(A,V)\bigr)\bigr|^{2}
      \hat{w}\bigr\} 
      = \int_{\mathbb{R}^{d}}\!\d{}x\; |w(x)|^{2}\!
      \int_{\mathbb{R}^{d}}\!\d{}y\; |f(x,y)|^{2}\,.\quad
    \end{equation}
    Here $\hat{w}$ denotes the bounded multiplication operator
    uniquely corresponding to $w$, and $\hat{w}^{*}$ denotes its
    Hilbert adjoint. 
  \end{nummer}
\end{theorem}

\begin{remarks}
  \begin{nummer}
  \item 
    The right-hand side of \eqref{fexeq} is well defined and continuous
    in $(x,y)\in\mathbb{R}^d\times\mathbb{R}^d$ by Lemma~\ref{kCarleman},
    Remark~\ref{Carleswap}, the boundedness of 
    $\e^{2tH(A,V)}F\bigl(H(A,V)\bigr)$ and the continuity of the 
    $\mathrm{L}^{2}(\mathbb{R}^{d})$-scalar product $\langle\cdot,
    \cdot\rangle$. Moreover, \eqref{fexeq} is independent of the
    chosen $t\in]0,\tau /2[$. 
  \item 
    The proof of Theorem~\ref{intkernel} is given in
    Section~\ref{intkernelproof} and rests on a more general result, which
    is formulated as Lemma~\ref{TBT}. This lemma is in the spirit
    of Thm.~B.7.8 in \cite{Sim82}, but, among others, we have relaxed
    a boundedness assumption in a suitable way.  
    Theorem~\ref{intkernel} itself may be viewed as a generalization
    of Thm.~B.7.1(d) in \cite{Sim82} from Kato-decomposable scalar
    potentials to ones with property
    \ass{V} and to vector potentials with property \ass{A}. But,
    whereas Thm.~B.7.1(d) in 
    \cite{Sim82} relies on resolvent techniques and requires the
    power-law decay $|F(E)| \le \mathrm{const.}( 1+|E|)^{-\alpha}$
    with $\alpha >d/2$ for
    energies $E$ in the spectrum of $H$, we work with the semigroup
    and thus need the decay property (\ref{Fcond}).
  \end{nummer}
\end{remarks}

\begin{corollary}
  \label{speckernel}
  Assume the situation of Theorem~\ref{intkernel} and let
  $I\subset\mathbb{R}$ be a Borel set in the real
  line which is bounded from above, $\sup I < \infty$. Then
  Theorem~\ref{intkernel} holds with $F= \Chi_I$, that is,  for the 
  spectral projection
  $\Chi_{I}\bigl(H(A,V)\bigr)$ associated with the energy regime $I$ 
  of the Schr{\"o}dinger operator $H(A,V)$. 
  Denoting the corresponding continuous integral kernel \eqref{fexeq}
  by $p_{I}$, Eq.\ \eqref{HStrace} takes the form
  \begin{equation}
    \label{tracechi}
    \mathrm{Trace}\bigl[\hat{w}^{*}\Chi_{I}\bigl(H(A,V)\bigr)
    \hat{w}\bigr] = \int_{\mathbb{R}^{d}}\!\d{}x\; |w(x)|^{2}\,
    p_{I}(x,x)
  \end{equation}
  for all $w\in\mathrm{L}^{\infty}_{\mathrm{G}}(\mathbb{R}^{d})$.
\end{corollary}

\begin{remark}
  The proof of Corollary~\ref{speckernel} is given in
  Section~\ref{intkernelproof}.
\end{remark}

Finally, we note that the functional calculus extends to integral 
kernels.

\begin{corollary}  
  \label{kernelcalculus}
  Assume the situation of Theorem~\ref{intkernel}. Then 
  \begin{equation}
    \label{fucakernel}
    f(x,y) = \int_{\mathbb{R}}\! \d p(E; x,y) \; F(E)
  \end{equation}
  holds for all $x,y\in\mathbb{R}^d$ and all $F$ obeying 
  \eqref{Fcond}. In addition, \eqref{fucakernel} holds  
  for the function $F$ given by $F(E) = \e^{-tE}$ with some arbitrary 
  $t \in ]0,\infty[$, in which 
  case one has to set $f=k_t$.   
  The right-hand side of 
  \eqref{fucakernel} is to be understood as a Lebesgue-Stieltjes 
  integral with respect to the complex ``distribution'' function 
  $\mathbb{R} \ni E \mapsto p(E;x,y) := 
  p_{]-\infty, E[}(x,y)$.
\end{corollary}

\begin{remark}
  The proof of Corollary~\ref{kernelcalculus} is given in
  Section~\ref{intkernelproof}.
\end{remark}

\subsection{Applications to random Schr{\"o}dinger operators}

The results of the previous subsection are nicely illustrated by
random Schr{\"o}\-dinger operators. In fact, certain random potentials
of wide-spread use in the physics literature on disordered systems
lead to Schr{\"o}dinger operators which are almost surely unbounded
from below and hence to Schr\"odinger semigroups which are almost
surely unbounded from above.

\begin{definition}
  A \emph{random scalar potential} $V$ on $\mathbb{R}^{d}$ is a random field 
  $V:  \Omega\times\mathbb{R}^{d}\to\mathbb{R}$, $(\omega,x)\mapsto 
  V^{(\omega)}(x)$, on a complete probability space
  $(\Omega, \mathcal{A},\mathbb{P})$ which is measurable with respect
  to the product of the sigma-algebra $\mathcal{A}$ of event sets in
  $\Omega$ and the sigma-algebra of Borel sets in $\mathbb{R}^{d}$. 
  Furthermore, a random scalar potential $V$ is said to satisfy property
  \begin{indentnummer}
  \item[\ass{S}] if there exist two reals 
    $ p_1 > p(d) $ and $ p_2 > p_1 d / \left[2 ( p_1 - p(d) ) \right]$ 
    such that 
    \begin{equation}\label{ess}
      \sup_{x \in \mathbb{Z}^d} \, \mathbb{E} \big[
     \|V\Chi_{\Lambda_1(x)}\|_{p_1}^{p_2}  \big]
      < \infty.
    \end{equation}
    Here, $\mathbb{E}[X]:=\int_\Omega \mathbb{P}(\d\omega)\, X^{(\omega)}$
    denotes the expectation of a (complex-valued) 
    random variable $X$ on $\Omega$, 
    and the real $ p(d) $ is defined as follows:
    $p(d) := 2$ if $d \leq 3 $, $ p(d) := d/2$ if $d \geq 5$ and $ p(4) > 2$, 
    otherwise arbitrary.
  \item[\ass{E}] if it is $\mathbb{R}^{d}$-ergodic with respect to the
    group of translations in $\mathbb{R}^{d}$, see \cite{Kir89}. 
  \item[\ass{I}] if
    \begin{equation}
      \sup_{x \in \mathbb{Z}^d} \, \mathbb{E} \big[
      \|V \Chi_{\Lambda_1(x)}\|^{2\vartheta+1}_{2\vartheta+1} \big] < \infty,
    \end{equation}
    where $\vartheta \in \mathbb{N}$ is the smallest integer with $
    \vartheta > d/4$. 
  \item[\ass{L}] if the finiteness condition
    \begin{equation}
      \mathcal{L}_{t} := \esssup_{x\in\mathbb{R}^{d}} \mathbb{E}
      \bigl[ \e^{-tV(x)}\bigr] < \infty
    \end{equation}
    holds for all $t>0$.
  \item[\ass{G}] if $V$ is a Gaussian random field \cite{Adl81,Lif95}
    which is $ \mathbb{R}^d $-homogeneous, has 
    zero mean, $ \mathbb{E}\left[\,V(0)\right] = 0 $, 
    and a covariance function
    $ x \mapsto C(x):= \mathbb{E}\left[\, V(x) V(0) \right] $ 
    that is continuous at the origin where it obeys 
    $ 0 < C(0) < \infty $.
    
  \end{indentnummer}
\end{definition}

\begin{remarks}
  \begin{nummer}
  \item 
    While property \ass{S} will assure the applicability of the results in the 
    previous subsection, property \ass{I}, respectively \ass{L}, is  mainly
    a technical one needed for the existence of the integrated density
    of states in Proposition~\ref{dosdef} below, respectively for the
    existence of the disorder-averaged semigroup in Corollary~\ref{avsemi}
    below. 
  \item
    Given \ass{E}, property \ass{I} simplifies to
    $\mathbb{E}\bigl[|V(0)|^{2\vartheta +1}\bigr] < \infty$ and
    property \ass{L} to $\mathcal{L}_{t} = \mathbb{E}
    \bigl[ \e^{-tV(0)}\bigr] < \infty$. Property \ass{L} implies
    neither \ass{S} nor \ass{I} and vice versa. Moreover, if
    $ d \neq 4 $, property~\ass{I} in general does
    not imply property~\ass{S}, even if property~\ass{E} is supposed.
    Given \ass{E}, a simple sufficient criterion for both \ass{S} and 
    \ass{I} to hold is the finiteness 
    \begin{equation}
      \mathbb{E}\bigl[|V(0)|^{p}\bigr] < \infty
    \end{equation}
    of the $p$-th absolute moment
    for some real $ p > \max\{3,d +1\} $.
    To prove this claim for property \ass{S},
    we choose $p_1 = p_2 = p $ in (\ref{ess}).
    For \ass{I} the claim follows from $ 2\vartheta \leq \max\{2,d\}$.   
  \item \label{gaussimply}
    If $V$ has property \ass{G}, then the standard
    Gaussian identity  
    \begin{equation}
      \label{gausschar}
      \mathbb{E}\biggl[\exp\biggl\{ \int_{\mathbb{R}^{d}}\!\zeta(\d
      x)\; V(x)  \biggr\}\biggr] = \exp\biggl\{ \frac{1}{2}\;
      \int_{\mathbb{R}^{d}}\!\zeta(\d x) \int_{\mathbb{R}^{d}}\!\zeta(\d
      y) \; C(x-y) \biggr\} \,. \qquad
    \end{equation}
    holds for all (finite) complex Borel measures $\zeta$ on $\mathbb{R}^{d}$.
    Accordingly, property \ass{G} implies properties \ass{S}, \ass{I} and
    \ass{L}, see Remark~3.9(iii) in \cite{HuLeMu01} for
    details. It also implies property \ass{E}, if the covariance function
    $C$ decays at infinity.   
  \end{nummer}
\end{remarks}
In order to apply the results of the previous subsection we need the
following 

\begin{lemma}
  \label{rpot}
  Let $V$ be a random scalar potential with property \ass{S}. Then  
  for $\mathbb{P}$-almost every $\omega\in\Omega$ the 
  realization $V^{(\omega)}: \mathbb{R}^d \rightarrow\mathbb{R}$, 
  $x\mapsto V^{(\omega)}(x)$ is a scalar potential with property \ass{V}. 
\end{lemma}

\begin{remark}
  The proof of the lemma is given in Section \ref{rpotproof}.
\end{remark}

For a vector potential with property \ass{A} and a random scalar
potential with property \ass{S} we thus infer from
Proposition~\ref{Hesssa} and Definition~\ref{Hdef} 
the existence of the \emph{random (magnetic)
  Schr{\"o}dinger operator} $H(A,V)$ given by the realizations
$H(A,V^{(\omega)})$, which are essentially self-adjoint on
$\mathcal{C}_{0}^{\infty}(\mathbb{R}^{d})$ for
$\mathbb{P}$-almost all $\omega\in\Omega$.

As an obvious consequence of Lemma~\ref{rpot} we note 

\begin{corollary}
  \label{rcor}
  Let $A$ be a vector potential with property \ass{A} and let $V$ be a random
  scalar potential with property \ass{S}. Then the results of
  Lemma~\ref{bridgelemma}, Theorem~\ref{fki}, Theorem~\ref{fkigroup}, 
  Theorem~\ref{intkernel}, Corollary~\ref{speckernel} and 
  Corollary~\ref{kernelcalculus}
  apply for $\mathbb{P}$-almost every $\omega\in\Omega$ to the
  realization $H(A,V^{(\omega)})$ of the random Schr{\"o}dinger
  operator as given by Definition~\ref{Hdef}.  
\end{corollary}

Corollary \ref{rcor} is the basis for the rigorous derivations of two
frequently used relations in the physics literature on disordered systems.

\subsubsection{Integrated density of states}

The first of these two relations is an integral-kernel representation
of the integrated 
density of states of random Schr{\"o}dinger operators. To formulate this
representation, we first recall one possible definition of
the integrated density of states in

\begin{proposition}
  \label{dosdef}
  Let $A$ be a vector potential with property \ass{C} and let $V$ be a random
  scalar potential with properties \ass{S}, \ass{E} and \ass{I}. 
  Let $\Gamma\subset \mathbb{R}^{d}$ be a bounded open cube and let 
  $\hat{\Chi}_{\Gamma}$ denote the bounded multiplication operator
  associated with the indicator function of $\Gamma$.
  Then the expectation value
  \begin{equation}
    \label{idos}
    N(E) := \frac{1}{|\Gamma|}\;
    \mathbb{E}\Bigl\{\mathrm{Trace} \Bigl[ \hat{\Chi}_{\Gamma} 
    \, \Chi_{]-\infty, E[} \bigl( H(A,V)\bigr) 
    \hat{\Chi}_{\Gamma}\Bigr]\Bigr\} 
  \end{equation}
  is well defined for every energy $E\in\mathbb{R}$ in terms of the
  spatially localized spectral projection associated with
  the half-line $]-\infty, E[$ of the random
  Schr{\"o}dinger operator $H(A,V)$. Furthermore it is independent of 
  $\Gamma$. The \emph{integrated density of states} $E \mapsto N(E)$ is  
  the unbounded left-continuous distribution function 
  of a positive Borel measure on the real line $\mathbb{R}\,$. 
\end{proposition}

\begin{proof}
  We refer to Thm.\ 3.1 in
  \cite{HuLeMu01} for the case $d \geq 2$ and to Thm.~5.20 in
  \cite{PaFi92} for the case $d=1$. 
\end{proof}

\begin{remark}
  Mostly, $N(E)$ is defined as
  the almost surely non-random quantity arising in the infinite-volume
  limit from the 
  number of eigenvalues per volume (counting multiplicities) of a
  finite-volume restriction of $H(A,V^{(\omega)})$ below $E$. This
  definition coincides with the one in Proposition~\ref{dosdef} above,
  as is shown in Cor.~3.3 of \cite{HuLeMu01} under the present
  assumptions on $A$ and $V$.
\end{remark}

On account of Corollary \ref{rcor} and \eqref{idos} we conclude

\begin{corollary}
  \label{doscor}
  Let $A$ be a vector potential with property \ass{C} and let $V$ be a random
  scalar potential with properties \ass{S}, \ass{E} and \ass{I}. Then
  the equality  
  \begin{equation}
    \label{dos=dos}
    N(E) = \mathbb{E} \bigl[ p(E;0,0)\bigr]
  \end{equation}
  holds for all $E\in\mathbb{R}\,$, where $p^{(\omega)}(E;\cdot,\cdot) 
  = p^{(\omega)}_{]-\infty, E[}$ denotes the continuous
  integral kernel of the spectral projection
  $\Chi_{]-\infty, E[} \bigl( H(A,V^{(\omega)})\bigr)$. 
  We recall that $p^{(\omega)}(E;\cdot,\cdot)$ exists
  for $\mathbb{P}$-almost all $\omega\in\Omega$ according to
  Corollary~\ref{rcor}.  
\end{corollary}

\begin{remarks}  
  \begin{nummer}
  \item 
    The corollary is proven in Section~\ref{rpotproof}.
  \item 
    The representation (\ref{dos=dos}) for the integrated
    density of states has been known previously from a rigorous point
    of view only under additional assumptions on the random scalar
    potential. For example, Remark~VI.1.5 in \cite{CaLa90} and 
    Remark~3.4 in \cite{HuLeMu01} require from the outset 
    the $\mathbb{P}$-almost sure 
    existence of continuous integral kernels for the spectral projections.
    A sufficient criterion for this requirement is that $V$ is 
    $\mathbb{P}$-almost surely Kato decomposable \cite{Sim82,BrHuLe00}.
    Earlier derivations of the representation \eqref{dos=dos} 
    by different authors require even stronger conditions on $V$, see 
    Thms.~5.18 and 5.23 in \cite{PaFi92}. The latter theorem, however, covers 
    differential operators more general than Schr\"odinger operators.
  \item To our knowledge, Corollary~\ref{doscor} provides the first
    rigorous derivation of the representation (\ref{dos=dos}) for a wide
    class of random scalar potentials. As we have seen, this class includes
    also random potentials leading to Schr\"odinger operators which are
    $\mathbb{P}$-almost surely unbounded from
    below. For example, this is the case if $V$ has
    properties \ass{G} and \ass{E} \cite{Kir89,CaLa90,PaFi92}.
    For such a choice of $V$ the relation \eqref{dos=dos} is
    frequently taken for granted in the physics literature on 
    disordered systems, see e.g.\ \cite{ShEf84,LeWo89,Efe97}.
  \item Corollary~\ref{doscor} strengthens Cor.~3.3 in \cite{HuLeMu01}
    in the sense that Eq.\ (3.6) in \cite{HuLeMu01} may be replaced by 
    Eq.\ (3.7) in \cite{HuLeMu01} without an additional assumption.
  \end{nummer}
\end{remarks}

\subsubsection{Disorder-averaged semigroup} 
The second application, for which Corollary~\ref{rcor} provides a rigorous
justification, concerns, loosely speaking, the expectation value of
the random operator exponential $\e^{-tH(A,V)}$.

\begin{corollary}
  \label{avsemi}
  Let $A$ be a vector potential with property \ass{A} and let $V$ be a random
  scalar potential with properties \ass{S} and \ass{L}. Moreover, let
  $t>0$ and let $k_{t}^{(\omega)}$ denote the continuous integral kernel 
  of $\e^{-t H(A,V^{(\omega)})}$. We recall that $k_{t}^{(\omega)}$ exists
  for $\mathbb{P}$-almost all $\omega\in\Omega$ according to
  Corollary~\ref{rcor}. Then 
  \begin{nummer}
  \item 
    the disorder-averaged integral kernel 
    $\overline{k_{t}} : \mathbb{R}^{d}\times\mathbb{R}^{d} \rightarrow
    \mathbb{C}\,$, 
    $(x,y)\mapsto \overline{k_{t}}(x,y) :=
    \mathbb{E}[k_{t}(x,y)]$ is well defined, 
    Hermitian in the sense that 
    $\overline{k_t}(x,y) = \overline{k_t}^{\;*}(y,x)$ for all
    $x,y\in\mathbb{R}^d$, continuous and dominated by the free heat kernel 
    according to    
    \begin{equation}
      \label{ueberfluessig}
      |\overline{k_{t}}(x,y)|  \le \mathcal{L}_{t} \; 
      \frac{\e^{-|x-y|^{2}/(2t)}}{(2\pi t)^{d/2}} 
    \end{equation}
    for all $x,y\in\mathbb{R}^{d}$. In particular, $\overline{k_{t}}(x,\cdot)
    \in \mathrm{L}^{\infty}_{\mathrm{G}}(\mathbb{R}^{d})$ for all $x \in
    \mathbb{R}^{d}$.     
    The mapping $\mathbb{R}^{d}\to\mathrm{L}^{2}(\mathbb{R}^{d})$, 
    $x \mapsto \overline{k_t}(x,\cdot)$ is strongly continuous.
  \item 
    the function $\overline{k_t}$ induces a bounded, self-adjoint and 
    positive Carleman 
    operator $T_{t}$ on $\mathrm{L}^{2}(\mathbb{R}^{d})$ in the sense that 
    \begin{equation}
      \label{uteq}
      T_{t}\psi := \int_{\mathbb{R}^{d}}\d\!y\; \overline{k_{t}}(\cdot,
      y)\, \psi(y) 
    \end{equation}
    for all $\psi\in \mathrm{L}^{2}(\mathbb{R}^{d})$ and that 
    $\overline{k_{t}}$ has the Carleman property \eqref{carledef}.
  \item  
    the image $T_{t}\psi$ of any
    $\psi\in\mathrm{L}^{2}(\mathbb{R}^{d})$ has a continuous
    representative in $\mathrm{L}^{2}(\mathbb{R}^{d})$ given by the 
    right-hand side of \eqref{uteq}. If even
    $\psi\in \mathrm{L}^{2}_{\mathrm{G}} (\mathbb{R}^{d})$, then one has in
    addition 
    $T_{t}\psi \in\mathrm{L}^{\infty}_{\mathrm{G}}(\mathbb{R}^{d})$ 
    and the equality
    \begin{equation}
      \label{Uinterpr}
      T_{t}\psi = \mathbb{E} \bigl[ \e^{-tH(A,V)}\psi \bigr]
    \end{equation}
    holds.
  \end{nummer}
\end{corollary}

\begin{remarks}
  \begin{nummer}
  \item
    The corollary is proven in Section~\ref{rpotproof}.
  \item In view of the equality in \eqref{Uinterpr}, the operator
    $T_{t}$ may be called the averaged semigroup 
    (operator). One should note, however, that
    the one-parameter family $\{T_{t}\}_{t \ge 0}$ is not a 
    semigroup in general.
  \item Assuming also properties \ass{C} and \ass{E}, the diagonal 
    of the kernel $\overline{k_t}$ is constant and given by the (two-sided)
    Laplace transform
    \begin{equation}
      \label{ktlaplace}
      \overline{k_t}(0,0) = \int_{\mathbb{R}}\!\d N(E)\; \e^{-tE}
    \end{equation}
    of the integrated density of states. This follows from Lemma~\ref{pmeas}
    below, Corollary~\ref{kernelcalculus}, integration by parts and Fubini's
    theorem. The latter two steps rely both on Lemma~\ref{kernelineq} below. 
  \end{nummer}
\end{remarks}

The content of Corollary~\ref{avsemi} is often used
in the physics literature on disordered solids and random media for
the special case 
where $V$ is a homogeneous Gaussian random potential, that is a random
scalar potential with property \ass{G}. For this choice of $V$, the
random Schr\"odinger operator $H(A,V)$ is $\mathbb{P}$-almost surely
unbounded from below \cite{Kir89,CaLa90,PaFi92}, but complies with the
assumptions of Corollary~\ref{avsemi} according to
Remark~\ref{gaussimply}. The corresponding Carleman kernel
$\overline{k_{t}}$ in Corollary~\ref{avsemi} can then be made more explicit by
applying Fubini's theorem and the standard Gaussian identity
\eqref{gausschar} with the finite measure $\zeta$ on $\mathbb{R}^d$ 
defined for $\mu_{x,y}^{0,t}$-almost every Brownian-bridge path $b$ by
its sojourn times $\zeta(\Lambda) := \int_{0}^{t}\d s\, 
\Chi_{\Lambda}(b(s))$ in Borel sets $\Lambda\subseteq \mathbb{R}^{d}$.
This leads to

\begin{corollary}
  \label{gauss}
  Let $A$ be a vector potential with property \ass{A} and let $V$ be a random
  scalar potential with property \ass{G}. Finally, let $t>0$. Then the
  assertions of Corollary~\ref{avsemi} hold with 
  \begin{eqnarray}
    \overline{k_{t}}(x,y) = \frac{\e^{-|x-y|^{2}/(2t)}}{(2\pi t)^{d/2}} &&
    \!\!\int\!\mu_{x,y}^{0,t}(\d b) \; e^{-S_{t}(A,0;b)} \nonumber\\*
    && \hspace*{-3em} \times
    \exp\biggl\{ \frac{1}{2}\; \int_{0}^{t}\!\d s \int_{0}^{t}\!\d s'
    \; C\bigl( b(s) - b(s')\bigr)\biggr\} 
    \label{gaussaverage}
  \end{eqnarray}
  for all $x,y\in\mathbb{R}^{d}$.
\end{corollary}

\begin{remark}
  The integral kernel \eqref{gaussaverage} obeys the inequality 
  \begin{equation}
    \label{siegfried}
    |\overline{k_{t}}(x,y)| \le \e^{-|x-y|^{2}/(2t)}
    \;\,\overline{k_{t}}(0,0)\,\big|_{A=0} \,,
  \end{equation}
  which is sharper, but less explicit than the estimate \eqref{ueberfluessig},
  when particularized to a Gaussian random potential.
  As to the validity of \eqref{siegfried} we note
  that by the diamagnetic inequality it suffices to consider 
  the situation with $A=0$. The latter was treated in
  \cite{LeWo89} by adapting an argument in the proof of Lemma~3.4 in
  \cite{DoVa83}. 
\end{remark}


%
\section{Proof of Lemma \protect{\lowercase{\ref{bridgelemma}}}}
\label{bridgeproof}

This section contains the probabilistic arguments which enter 
Lemma~\ref{bridgelemma}. 

\begin{proof}[of Lemma \ref{bridgelemma}]
  To begin with, we establish the bound (\ref{kbound}). In so doing we
  also show that the Brownian-bridge functional $b \mapsto \exp\{
  -S_{t}(A,V;b)\}$ is $\mu_{x,y}^{0,t}$-integrable and hence
  (\ref{kern}) well defined. To this end, we successively apply the
  triangle and the Cauchy-Schwarz inequality to the (absolute square
  of the) Brownian-bridge expectation in (\ref{kern})
\begin{eqnarray}
  \left| \int\mu_{x,y}^{0,t}(\d b)\; \e^{-S_{t}(A,V;b)} 
  \right|^{2}
  && \!\!\! \le \left(\int\mu_{x,y}^{0,t}(\d b)\;  | \e^{-S_{t}(A,V;b)} | 
  \right)^{2}\nonumber\\
  && \hspace*{-3em} = \left(\int\mu_{x,y}^{0,t}(\d b)\;   \e^{-S_{t}(0,V;b)}  
  \right)^{2}\nonumber\\
  && \hspace*{-3em} \le \int\mu_{x,y}^{0,t}(\d b)\; \e^{-S_{t}(0,2V_{1};b)}
  \int\mu_{x,y}^{0,t}(\d b)\; \e^{-S_{t}(0,2V_{2};b)}  \,. 
  \nonumber\\*
  \label{kboundstart}         
\end{eqnarray} 
It follows from Eq.~(1.3.5) in \cite{Szn98} that
\begin{equation} 
  \label{secondexpectation}
  \int\mu_{x,y}^{0,t}(\d b)\; \e^{-S_{t}(0,2V_{1};b)} \le 
  C_{0}(t)\, \exp\left\{ {|x-y|^{2}}/{(4t)}\right\} 
\end{equation}
thanks to $V_{1}\in\mathcal{K}_{\pm}(\mathbb{R}^{d})$ by property
\ass{V}. Here $C_{0}(t)$ is strictly positive and continuous in
$t\in ]0,\infty[$. Moreover, it is independent of $x,y \in
\mathbb{R}^d$. As to the second expectation in the last line of
(\ref{kboundstart}), the inequality (\ref{Wcond}) and the proof of
Thm.~2.1 in \cite{Sim00} give for all $\lambda>0$ and all
$\varepsilon\in ]0,(\lambda t^{2})^{-1}[$ the estimate
\begin{eqnarray}
  \int\mu_{x,y}^{0,t}(\d b)\; \e^{-S_{t}(0,\lambda V_{2};b)}  & \le & 
  \int\mu_{x,y}^{0,t}(\d b)\; \e^{S_{t}(0,\lambda |V_{2}|;b)} \nonumber\\
  & \le & 
  \Upsilon(\lambda\varepsilon t^2)\, \e^{\lambda t v_\varepsilon} \,
  \e^{2\lambda\varepsilon t (|x|^2 +|y|^2)}\,,
  \label{simonbound}
\end{eqnarray}
where $\Upsilon(\xi):= \int_0^1\d\sigma\, [1-4\xi\sigma(1-\sigma)]^{-d/2}$
is increasing in $\xi$ and finite for all $\xi\in[0,1[$. 
Together with (\ref{secondexpectation}) and
(\ref{kboundstart}), the estimate \eqref{simonbound} with $\lambda=2$
establishes (\ref{kbound}) for all $\delta \in ]0,t^{-1}[$ by
identifying $\delta$ with $2\varepsilon t$. For arbitrary $\delta \ge t^{-1}$
the estimate \eqref{kbound} then follows from the monotonicity of 
$\delta \mapsto \e^{\delta |x|^2 + \delta |y|^2}$.

Next we prove the properties of $k_t$ claimed in part \itemref{kerneldef} 
of the lemma. 
The Hermiticity and the semigroup property of
$k_{t}$ are a consequence of the time-reversal
invariance and the Markov property of the Brownian bridge,
respectively. This follows from the line of reasoning in the proof of
Eqs.~(1.3.6) and (1.3.7) in \cite{Szn98}. For the proof of the
continuity of $k_{t}$ we refer to Corollary~\ref{contcor1} below.

Finally, we turn to the proof of part \itemref{kCarleman}. 
The claim \eqref{kstark} is immediate from the estimate \eqref{kbound}. 
The semigroup property \eqref{markov} and the 
Hermiticity give
\begin{equation}
  \| k_{t}(x,\cdot) - k_{t}(z,\cdot) \|^2_{2}
  = k_{{2t}}(x,x) - k_{{2t}}(z,x) - k_{{2t}}(x,z) + k_{{2t}}(z,z)\quad
\end{equation}
for all $x,z \in\mathbb{R}^{d}$. This equality together with the 
continuity of $k_{2t}$ establishes
the strong continuity of the mapping 
$\mathbb{R}^{d}\to\mathrm{L}^{2}(\mathbb{R}^{d})$, $x \mapsto 
k_{t} (x,\cdot)$.
\end{proof}

Lemma~\ref{unicont} below is our basic technical result for deducing
the already claimed continuity of $k_{t}$. 
It will also enter
the proof of the Feynman-Kac-It\^{o} formula in the next section. For
both purposes Lemma~\ref{unicont} will provide an 
approximation argument. We use it to deduce the desired properties  
from corresponding ones of Schr\"odinger semigroups with
regularized scalar potentials which are Kato decomposable.

\begin{definition}
  \label{vrdef}
  Given any real $R>0$ and a scalar potential $V$ with property 
  \ass{V}, we define a \emph{regularized
    scalar potential} $V_{R}\in \mathrm{L}^{2}_{\mathrm{loc}}(\mathbb{R}^{d})
  \cap \mathcal{K}_{\pm}(\mathbb{R}^{d})$ by setting
  \begin{equation}
    V_{R} := V_{1} + V_{2, R}\,,
  \end{equation}
  where its truncated part $x\mapsto V_{2, R}(x):= \Theta(R-|x|)\, V_{2}(x)$ 
  lies in $\mathrm{L}^\infty(\mathbb{R}^d)$.
\end{definition}

\begin{lemma}
  \label{unicont}
  Let $A$ be a vector potential with property \ass{A} and let $V$ be a scalar
  potential with property \ass{V}. For
  $t>0$, $R>1$ and $x,y\in\mathbb{R}^{d}$ define the regularized kernel 
  \begin{equation}
    k_{t}^{(R)}(x,y) := \frac{\e^{-|x-y|^{2}/(2t)}}{(2\pi t)^{d/2}}
    \int\!\mu_{x,y}^{0,t}(\d b)\;\e^{-S_{t}(A,V_{R};b)} \,.
  \end{equation}
  Then for every triple $\tau_1,\tau_2,\tilde{\rho} \in ]0,\infty[$ with 
  $\tau_1 \le \tau_2$ there exists $\rho \in ]0,\infty[$ such that one has the
  uniform-type-of convergence
  \begin{equation}
    \lim_{R\to\infty} \sup_{x,y\in\mathbb{R}^{d}} \sup_{t\in[\tau_{1},
      \tau_{2}]} \left[ \e^{\rho|x|^{2} -
        \tilde{\rho}|y|^{2}} \, | k_{t}(x,y) -
      k_{t}^{(R)}(x,y)| \right] = 0\,.
  \end{equation}
\end{lemma}

\begin{proof}
  Given a H\"older exponent $p\in]1,\infty[$, we denote by $p':=
  (1-p^{-1})^{-1}$ its conjugate exponent. Moreover, we let
  $t\in[\tau_{1},\tau_{2}]$ arbitrary. Then
  the triangle and the H\"older inequality yield
  \begin{eqnarray}
    \lefteqn{\left| \int\mu_{x,y}^{0,t}(\d b)\; \left[ \e^{-S_{t}(A,V;b)} -
          \e^{-S_{t}(A,V_{R};b)}  \right] \right| } \nonumber\\*
    && \le \int\mu_{x,y}^{0,t}(\d b)\; \e^{-S_{t}(0,V_{1};b)} \left| 
      \e^{-S_{t}(0,V_{2};b)} - \e^{-S_{t}(0,V_{2, R};b)} \right| 
    \nonumber\\*
    && \le \left[ \int\mu_{x,y}^{0,t}(\d b)\; \e^{-S_{t}(0,pV_{1};b)}
    \right]^{\frac{1}{p}} \left[ \int\mu_{x,y}^{0,t}(\d b)\,\left| 
        \e^{-S_{t}(0,V_{2};b)} - \e^{-S_{t}(0,V_{2, R};b)} \right|^{p'}
    \right]^{\frac{1}{p'}} .\nonumber\\*
    \label{contstart}
  \end{eqnarray}
  The first expectation in the last line of
  (\ref{contstart}) is bounded according to
  \begin{equation}
    \label{contsecond}
    \left[ \int\mu_{x,y}^{0,t}(\d b)\; \e^{-S_{t}(0,pV_{1};b)}
    \right]^{1/p} \le C_{1} \, \exp\left\{
      {|x-y|^{2}}/{(4\tau_{1}p)}\right\}\,,
  \end{equation}
  confer \eqref{secondexpectation}.
  Here $C_{1}\equiv C_{1}(p,\tau_{1},\tau_{2})$ is a finite constant.
  In order to bound the second expectation in the last line of
  (\ref{contstart}) we employ the elementary inequality $|\e^{r}
  -\e^{r'}| \le |r-r'| \, \e^{\max\{r,r'\}}$ for $r,r'\in\mathbb{R}$
  together with $|V_{2, R}| \le |V_{2}|$ and the Cauchy-Schwarz
  inequality. This gives
  \begin{eqnarray}
    \lefteqn{\int\mu_{x,y}^{0,t}(\d b)\,\left| 
        \e^{-S_{t}(0,V_{2};b)} - \e^{-S_{t}(0,V_{2, R};b)}
      \right|^{p'} } \nonumber\\*
    && \le \int\mu_{x,y}^{0,t}(\d b)\, \e^{S_{t}(0,p'|V_{2}|\,;b)}\,
    \bigl| S_{t}(0,V_{2}-V_{2, R};b) \bigr|^{p'} \nonumber\\*
    && \le \left[ \int\mu_{x,y}^{0,t}(\d b)\; \e^{S_{t}(0,
        2p'|V_{2}|\,;b)} \right]^{1/2}
    \left[ \int\mu_{x,y}^{0,t}(\d b)\, \bigl|
      S_{t}(0,V_{2}-V_{2, R};b) \bigr|^{2p'}  
    \right]^{1/2}.\nonumber\\*
    \label{contmiddle}
  \end{eqnarray}
  The first expectation in the last line of (\ref{contmiddle}) can be
  estimated as in (\ref{simonbound}), 
  \begin{equation}
    \label{contnext}
    \int\mu_{x,y}^{0,t}(\d b)\; \e^{S_{t}(0,
        2p'|V_{2}|\,;b)} \le C_{2}^{2p'} \,\exp\bigl\{
      4p' \varepsilon \tau_{2} \bigl(
      |x|^{2} + |y|^{2}\bigr)\bigr\}\,, \qquad
  \end{equation}
  where $\varepsilon \in ]0, (2p'\tau_{2}^2)^{-1}[$ is arbitrary and  
  $C_{2}\equiv C_{2}(p,\varepsilon,\tau_{2})$ is another finite constant. 
  Here we have used the monotonicity of the right-hand side of 
  \eqref{simonbound} in $t$.
  To bound the second expectation in the last line of
  (\ref{contmiddle}) we observe that 
  \begin{equation}
    \label{v2cons}
    |V_{2}(x) -V_{2, R}(x)| \le (\varepsilon |x|^{2} +v_\varepsilon) \,
    \Theta(|x|-R) \le 
    (\varepsilon + v_\varepsilon) \; \frac{|x|^4}{R^{2}} 
  \end{equation}
  for all $\varepsilon >0$ and Lebesgue-almost all
  $x\in\mathbb{R}^{d}$. Here we have exploited $R>1$ and the 
  ``Chebyshev'' inequality $\Theta(\xi -1) \le \xi^2$, $\xi\in\mathbb{R}$. 
  By the Jensen and the triangle
  inequality, Fubini's theorem  and upon standardizing the Brownian
  bridge according to 
  $b(s)=: t^{1/2} \, \tilde{b}(s/t) + x + (y-x)s/t$,
  the estimate \eqref{v2cons} yields
  \begin{eqnarray}
    \lefteqn{\int\mu_{x,y}^{0,t}(\d b)\, \bigl|
      S_{t}(0,V_{2}-V_{2, R};b) \bigr|^{2p'}}\nonumber\\*
    && \le \Bigl(\frac{(\varepsilon + v_\varepsilon)t}{R^{2}}\Bigr)^{2p'}
    \int_{0}^{t} \frac{\d s}{t} 
    \int\mu_{x,y}^{0,t}(\d b)\; |b(s)|^{8p'} \nonumber\\*
    && = \Bigl(\frac{(\varepsilon + v_\varepsilon)t}{R^{2}}\Bigr)^{2p'}
    \int_{0}^{1} \d\sigma\;\int\mu_{0,0}^{0,1} (\d\tilde{b})\; 
    |t^{1/2}\tilde{b}(\sigma) + x + (y-x)\sigma|^{8p'}.\nonumber\\*
  \end{eqnarray}
  This result and several applications of the elementary inequality
  \begin{equation}
    \label{elineq}
    |r+r'|^{\alpha} \le 2^{\alpha } \bigl( |r|^{\alpha} +
    |r'|^{\alpha}\bigr) 
  \end{equation}
  for $\alpha >0$ and $r,r'\in\mathbb{R}^{d}$
  show that there exist two further finite constants 
  $C_{3}\equiv C_{3}(p,\varepsilon)$ and  
  $C_{4}\equiv C_{4}(p,\varepsilon)$ such that 
  \begin{equation}
    \left[ \int\mu_{x,y}^{0,t}(\d b)\, \bigl|
      S_{t}(0,V_{2}-V_{2, R};b) \bigr|^{2p'}  
    \right]^{1/(2p')}  
   \le \frac{\tau_2}{R^2} \;\bigl[  C_{3}
   \tau_{2}^{2} +  C_{4} \bigl(|x|^4 + |y|^{4}\bigr) \bigr]\,. 
    \label{contfinal}
  \end{equation}
  Combining (\ref{contstart}), (\ref{contsecond}), (\ref{contmiddle}),
  (\ref{contnext}) and (\ref{contfinal}), we obtain
  \begin{eqnarray}
    \lefteqn{\left| \int\mu_{x,y}^{0,t}(\d b)\; \left[ \e^{-S_{t}(A,V;b)} -
          \e^{-S_{t}(A,V_{R};b)}  \right] \right|} \nonumber\\*
    && \le \frac{C_{1}C_{2}\tau_2}{R^{2}}\; \bigl[  C_{3} \tau_{2}^{2} +  
    C_{4} \bigl( |x|^{4} + |y|^{4} \bigr)\bigr]\exp\left\{ \frac{
      |x-y|^{2}}{4\tau_{1}p} + 2\varepsilon\tau_{2} \bigl(|x|^{2} 
    + |y|^{2}\bigr)\right\} \nonumber\\*
  \end{eqnarray}
  for all $t\in [\tau_1,\tau_2]$, all $\varepsilon \in ]0, 
  (2p'\tau_2^2)^{-1}[ $ and all $x,y\in\mathbb{R}^d$.
  Another application of \eqref{elineq} and choosing 
  $p=2\tau_{2}/\tau_{1} \ge 2$ then yields 
  \begin{eqnarray}
    \lefteqn{\sup_{t\in[\tau_{1},\tau_{2}]} 
      \left[ \e^{\rho|x|^{2} -
          \tilde{\rho}|y|^{2}} \, | k_{t}(x,y) -
        k_{t}^{(R)}(x,y)| \right] } \nonumber\\*
  && \le \frac{C_{1}C_{2}\tau_2}{R^{2}(2\pi\tau_{1})^{d/2}}\;
  \bigl[  C_{3} \tau_{2}^{2} +  C_{4} \bigl(  |x|^{4}
  + |y|^{4} \bigr)\bigr] \nonumber\\*
  && \phantom{\le} \times \exp\big\{ -\bigl[ 1/(4\tau_{2}) 
   -4 \rho  -8\varepsilon\tau_2 \bigr] |x-y|^{2} -  
  (\tilde{\rho} -4\rho - 10\, \varepsilon\tau_{2}) |y|^{2}
  \big\} \nonumber\\*
\end{eqnarray}
for all $\rho,\tilde{\rho} >0$, all $\varepsilon \in ]0,
(2\tau_{2}-\tau_{1})/(4\tau_{2}^{3})[$ and all $x,y\in\mathbb{R}^{d}$.
The assertion of the lemma now follows by choosing $\rho$ and 
$\varepsilon$ so small that $4\rho + 10\,
\varepsilon\tau_{2}  < \min\{\tilde{\rho},(4\tau_2)^{-1}\}$. 
\end{proof}

Lemma~\ref{unicont} possesses an immediate corollary, which 
completes the proof of Lemma~\ref{bridgelemma}.

\begin{corollary}
  \label{contcor1}
  The function
  \begin{equation}
    ]0,\infty[ \times \mathbb{R}^{d} \times\mathbb{R}^{d} \rightarrow
    \mathbb{C}, \qquad  (t,x,y)
    \mapsto k_{t}(x,y)
  \end{equation}
  is continuous under the assumptions of Lemma~\ref{bridgelemma}.
\end{corollary}

\begin{proof}
  Since by assumption $V_{R}$ lies in
  $\mathcal{K}_{\pm}(\mathbb{R}^{d})$ and both $|A|^{2}$ and
  $\nabla\cdot A$ lie in $\mathcal{K}_{\mathrm{loc}}(\mathbb{R}^{d})$,
  Thm.~6.1 in \cite{BrHuLe00} for the case $d\ge 2$, respectively
  Prop.~1.3.5 in \cite{Szn98} for the case $d=1$, guarantee the 
  continuity of the function
  \begin{equation}
    ]0,\infty[ \times \mathbb{R}^{d} \times\mathbb{R}^{d} \rightarrow
    \mathbb{C}, \qquad (t,x,y)
    \mapsto k_{t}^{(R)}(x,y) 
  \end{equation}
  for all $R>0$. But according to Lemma~\ref{unicont} the kernel
  $k_{\scriptscriptstyle\bullet}$ is the locally uniform limit of
  $k_{\scriptscriptstyle\bullet}^{(R)}$ as
  $R\to\infty$. Hence, $k_{\scriptscriptstyle\bullet}$ inherits the continuity
  properties of $k_{\scriptscriptstyle\bullet}^{(R)}$.
\end{proof}


%
\section{Proofs of Theorem \protect{\lowercase{\ref{fki}}} and 
Theorem~\protect{\lowercase{\ref{fkigroup}}}}  
\label{fkiproof}

Given the two probabilistic Lemmata~\ref{bridgelemma}
and~\ref{unicont}, the additional 
arguments needed to prove Theorem~\ref{fki} and Theorem~\ref{fkigroup}
are purely analytic. First, we exploit the fact that the
function $k_{t}$, as defined in Lemma~\ref{bridgelemma},
is a Carleman kernel \cite{Wei80}.

\begin{lemma}
  \label{kate}
  Let $A$ be a vector potential with property \ass{A} and let $V$ be a
  scalar potential with property \ass{V}. For $t>0$ we denote by $K_{t}$ 
  the integral operator induced by the kernel $k_{t}$ with domain
  \begin{equation}
    \dom(K_{t}) := \Bigl\{\psi\in\mathrm{L}^{2}(\mathbb{R}^{d}):
      \int_{\mathbb{R}^{d}}\d{}y\; k_{t}(\cdot,y)\,\psi(y) \in
      \mathrm{L}^{2}(\mathbb{R}^{d}) \Bigr\}
  \end{equation}
  and action
  \begin{equation}
    \label{Ktaction}
    K_{t}\psi := \int_{\mathbb{R}^{d}}\d{}y\; k_{t}(\cdot,y)\,\psi(y) 
  \end{equation}
  for all $\psi\in\dom(K_{t})$.
  Then $K_{t}$ is a maximal Carleman operator, hence closed, 
  and its domain is dense thanks to the inclusion
  \begin{equation}
    \label{domincl}
    \mathrm{L}^{2}_{\mathrm{G}}(\mathbb{R}^{d}) \subseteq
     \dom(K_{t})\,.
  \end{equation} 
  Moreover, the image $K_{t}\psi$ of any $\psi\in\dom(K_t)$
  has a continuous representative in $\mathrm{L}^{2}(\mathbb{R}^{d})$ 
  given by the right-hand side of \eqref{Ktaction}. If even
  $\psi\in\mathrm{L}^2_{\mathrm{G}}(\mathbb{R}^d)$, then, in addition,
  $K_t \psi \in\mathrm{L}^{\infty}_{\mathrm{G}}(\mathbb{R}^d)$. 
\end{lemma}

\begin{proof}[of Lemma~\ref{kate}]
By Lemma~\ref{kerneldef} and~\itemref{kCarleman} we know that 
$k_{t}$ is a Hermitian Carleman kernel.
Thus, Thm.~6.13(a) in \cite{Wei80} yields the closedness of the induced
maximal Carleman operator $K_{t}$. The inclusion (\ref{domincl}) is implied by
Remark~\ref{lepsincl} and the inclusion $K_{t}
\mathrm{L}_{\mathrm{G}}^{2}(\mathbb{R}^{d}) \subseteq
\mathrm{L}_{\mathrm{G}}^{\infty}(\mathbb{R}^{d})$, which we prove
next. To do so, we note that (\ref{kbound}) implies
\begin{equation}
  \label{basic}
  \sup_{x\in\mathbb{R}^{d}} \Bigl[\e^{\rho |x|^{2}}
  |k_{t}(x,y)| \Bigr] \le a_{t}^{(\delta)}\; \e^{(4\rho +5\delta) |y|^{2}}
\end{equation}
for all $\rho,\delta >0$ with $\rho + \delta < 1/(16t)$ and 
all $y\in\mathbb{R}^d$.
In deriving (\ref{basic}) we have also used the elementary inequality
\eqref{elineq} with $r=x-y$, $r'=y$ and $\alpha =2$. 

Consequently, given any $\psi\in\mathrm{L}_{\mathrm{G}}^{2}
(\mathbb{R}^{d})$, we get 
\begin{equation}
  \label{liesin}
  \esssup_{x\in\mathbb{R}^{d}} \Bigl|\e^{\rho |x|^{2}}
  (K_{t}\psi)(x) \Bigr|  \le
  a_{t}^{(\delta)} \int_{\mathbb{R}^{d}}\!\d{}y\;
  \e^{(4\rho+5\delta)|y|^{2}} \, |\psi(y)|\,.
\end{equation}
Now, choosing $\rho$ and $\delta$ small enough, the right-hand side of 
\eqref{liesin} is finite since $\mathrm{L}_{\mathrm{G}}^{2}(\mathbb{R}^{d})
\subseteq \mathrm{L}_{\mathrm{G}}^{1}(\mathbb{R}^{d})$ by
Remark~\ref{lepsincl}.

In order to complete the proof of the lemma we have to show the
continuity of $K_{t}\psi$ 
for all $\psi\in\dom(K_t)$. To this end we observe 
\begin{equation}
  \label{contest}
  \bigl| (K_{t}\psi)(x) - (K_{t}\psi)(x')\bigr| \le
  \Vert \psi\Vert_2 \; \Vert k_{t}(x,\cdot) -k_{t}(x',\cdot) \Vert_2
\end{equation}
by the triangle and the Cauchy-Schwarz inequality for all 
$x,x'\in\mathbb{R}^{d}$. 
The desired result now follows from the strong continuity of 
$x \mapsto k_t(x,\cdot)$ in Lemma~\ref{kCarleman}.
\end{proof}

We will eventually prove Theorem~\ref{fki} by showing
the operator equality $K_{t} = \e^{-tH(A,V)}$. As an initial step
we recall Definition~\ref{vrdef} and employ Lemma~\ref{unicont} in
order to establish strong convergence of 
the regularized operator exponentials $\e^{-tH(A,V_{R})}$ to $K_{t}$ on
$\mathrm{L}^{2}_{\mathrm{G}}(\mathbb{R}^{d})$ as $R\to\infty$.  

\begin{lemma}
  \label{strongcvg}
  Let $t>0$, $\psi\in\mathrm{L}^{2}_{\mathrm{G}}(\mathbb{R}^{d})$ and
  suppose the assumptions of Theorem~\ref{fki}. Then
  \begin{equation}
    \lim_{R\to\infty} \Vert \e^{-tH(A,V_{R})}\psi -K_{t}\psi \Vert_{2} =0
  \end{equation}
  holds.
\end{lemma}

\begin{proof}
  We recall from Thm.~6.1  in \cite{BrHuLe00} for the case $d\ge 2$,
  respectively  
  from Eq.~(6.6) in \cite{Sim79a} or from Eqs.~(1.3.3), (1.3.4) and 
  Exercise~1.4.2 in \cite{Szn98} for the case $d=1$,
  the Feynman-Kac-It\^{o}
  formula for the bounded semigroup with the regularized
  potential
  \begin{equation}
    \e^{-tH(A,V_{R})}\psi =\int_{\mathbb{R}^{d}}\!\d{}y\;
    k_{t}^{(R)}(\cdot,y) \, \psi(y) \,,
  \end{equation}
  valid for all $\psi\in\mathrm{L}^{2}(\mathbb{R}^{d})$. Now, 
  given any $\psi\in\mathrm{L}^{2}_{\mathrm{G}}(\mathbb{R}^{d})$ there exists 
  $\tilde{\rho} >0$ such that $\Vert \e^{\tilde{\rho}|\cdot|^2}\psi\Vert_1
  < \infty$ by Remark~\ref{lepsincl}. 
  Lemma~\ref{unicont} then yields the existence of $\rho >0$
  such that the right-hand side of the estimate
  \begin{eqnarray}
    \Vert \e^{-tH(A,V_{R})}\psi -K_{t}\psi \Vert_{2}^{2} 
     &=& \int_{\mathbb{R}^{d}}\!\d{}x \left|
        \int_{\mathbb{R}^{d}}\!\d{}y\; \bigl[ k_{t}^{(R)}(x,y) -
        k_{t}(x,y) \bigr] \,\psi(y)\right|^{2}  \nonumber\\*
    &\le& \int_{\mathbb{R}^{d}}\!\d{}x \;
    \e^{-2\rho|x|^{2}} \bigg[
    \int_{\mathbb{R}^{d}}\!\d{}y\;
    \e^{\tilde{\rho}|y|^{2}} \,|\psi(y)| \nonumber\\*
    &&\hspace*{1.2cm} \times \;\e^{\rho|x|^{2} -
      \tilde{\rho}|y|^{2}}\, \bigl| k_{t}^{(R)}(x,y) -
        k_{t}(x,y) \bigr| \bigg]^{2} \nonumber\\
    &\le& \bigg[ \sup_{x,y\in\mathbb{R}^{d}} \Big(
    \e^{\rho|x|^{2} -  \tilde{\rho}|y|^{2}}\,
    \bigl| k_{t}^{(R)}(x,y) -  k_{t}(x,y) \bigr|\Big)\bigg]^{2}
    \nonumber\\*
    &&  \times \;[\pi/(2\rho)]^{d/2}\;
     \Vert\e^{\tilde{\rho}|\cdot|^{2}} \,\psi\Vert_1^2 
  \end{eqnarray}
  vanishes as $R \to\infty$.
\end{proof}

\begin{remark}
  One can even show that the convergence in Lemma~\ref{strongcvg}
  holds with respect to the $\mathrm{L}^{p}(\mathbb{R}^{d})$-norm for
  arbitrary $p\in[1,\infty]$, if one
  requires $\psi\in\mathrm{L}^{p}_{\mathrm{G}}(\mathbb{R}^{d})$, see
  also Remark~\ref{lpwell}.
\end{remark}

The next lemma concerns a certain stability of
strong-resolvent convergence. It will be the basis for an
argument similar to the one provided by Thm.~3.1 in \cite{Sim00}. 

\begin{lemma}
  \label{src}
  For $n\in\mathbb{N}$ let $A_{n}$ and $A$ be self-adjoint operators
  acting on a complex Hilbert space and let $G: \mathbb{R}
  \rightarrow\mathbb{R}$ be a continuous function. Define $G(A_{n})$
  for $n\in\mathbb{N}$ and $G(A)$ via the spectral theorem and the
  functional calculus as self-adjoint operators. Then strong-resolvent
  convergence of $A_{n}$ to $A$ as $n\to\infty$ implies
  strong-resolvent convergence of $G(A_{n})$ to $G(A)$.
\end{lemma}

\begin{proof}
  For $z\in\mathbb{C}$ with $\Im z \neq 0$ we define the bounded
  continuous function $R_{z}: \mathbb{R}\rightarrow\mathbb{C}$,
  $\lambda\mapsto R_{z}(\lambda):= (\lambda-z)^{-1}$. Hence, the
  composition $R_{z}\comp G$ is also a bounded and continuous function
  on $\mathbb{R}$. Therefore, $(R_{z}\comp G)(A_{n}) =
  R_{z}(G(A_{n}))$ converges strongly to $(R_{z}\comp G)(A) =
  R_{z}(G(A))$ as $n\to\infty$ by Thm.~VIII.20(b) in \cite{ReSi80} or
  Thm.~9.17 in \cite{Wei80}.
\end{proof}

Having these auxiliary results at our disposal, we can proceed to prove
-- as an intermediate step -- Theorem~\ref{fkiintegral}, which is
analogous to the claim of Remark~1 after Thm.~1.2 in \cite{Sim00}.

\begin{lemma}
  \label{halfway}
  Let $t>0$. Under the assumptions of Theorem~\ref{fki} one has 
  $\mathrm{L}^2_{\mathrm{G}}(\mathbb{R}^d) \subseteq
  \dom\bigl(\e^{-tH(A,V)}\bigr)$ and the Feynman-Kac-It\^{o} formula 
  \begin{equation}
    \label{fkieps}
    \e^{-tH(A,V)}\psi = K_{t}\psi
  \end{equation}
  holds for all $\psi\in\mathrm{L}^2_{\mathrm{G}}(\mathbb{R}^d)$.
  In particular, $\e^{-tH(A,V)}$ and thus $K_t$ are both symmetric on 
  $\mathrm{L}^2_{\mathrm{G}}(\mathbb{R}^d)$. 
\end{lemma}

\begin{proof}[of Lemma~\ref{halfway}]
  The Schr\"odinger operators $H(A,V)$ and $H(A,V_{R})$,
  $R>0$, are all essentially self-adjoint on
  $\mathcal{C}_{0}^{\infty}(\mathbb{R}^{d})$ according to
  Proposition~\ref{Hesssa} and Definition~\ref{Hdef}. Moreover,
  $H(A,V_{R})$ converges strongly to 
  $H(A,V)$ on $\mathcal{C}_{0}^{\infty}(\mathbb{R}^{d})$ as
  $R\to\infty$. This can be inferred from \eqref{Wcond} and the estimate
  \begin{eqnarray}
    \Vert H(A,V_{R})\varphi -H(A,V)\varphi \Vert_{2}^{2} &=&
      \int_{\mathbb{R}^{d}}\!\d{}x\; \bigl|V_{2}^{(R)}(x) -
      V_{2}(x)\bigr|^{2}\, |\varphi(x)|^{2}  \nonumber\\*
    &\le&  \int_{\mathbb{R}^{d}}\!\d{}x\; \Theta(|x| - R)\, 
    \bigl( \varepsilon |x|^{2} + v_\varepsilon\bigr)^{2}\,
    |\varphi(x)|^{2}\,,   \nonumber\\*
    \label{strongcvg2}
  \end{eqnarray}
  which is valid for all $\varepsilon >0$ and all
  $\varphi\in\mathcal{C}_{0}^{\infty}(\mathbb{R}^{d})$. 
  The
  right-hand side of \eqref{strongcvg2} vanishes, if $R$ is large enough.
  Therefore,
  Thm.~VIII.25(a) in \cite{ReSi80} implies that $H(A,V_{R})$ converges
  to $H(A,V)$ in strong-resolvent sense as $R\to\infty$, and thus, thanks
  to Lemma~\ref{src}, $\e^{-tH(A,V_{R})}$ converges to $\e^{-tH(A,V)}$
  as $R\to\infty$ in strong-resolvent sense for all $t>0$. Since
  the operators $\e^{-tH(A,V_{R})}$  and $\e^{-tH(A,V)}$ are self-adjoint,
  strong-resolvent convergence is equivalent to $\e^{-tH(A,V)}$ being
  the strong-graph 
  limit of $\e^{-tH(A,V_{R})}$ as $R\to\infty$ by Thm.~VIII.26 in
  \cite{ReSi80}. Thus, by definition of this limit, the graph 
  \begin{eqnarray}
    &\mathcal{G}_{t} := \bigl\{ (\psi,\phi)\in
    \mathrm{L}^{2}(\mathbb{R}^{d}) \times
    \mathrm{L}^{2}(\mathbb{R}^{d}) : \psi\in
    \dom\bigl(\e^{-tH(A,V)}\bigr),
    \phi=\e^{-tH(A,V)}\psi\bigr\} &\nonumber\\*
  \end{eqnarray}
  of $\e^{-tH(A,V)}$ consists of all pairs $(\psi,\phi)\in
  \mathrm{L}^{2}(\mathbb{R}^{d}) \times \mathrm{L}^{2}(\mathbb{R}^{d})$
  for which there exists a sequence $\{\psi_{R}\}_{R}$ with $\psi_{R} \in
  \dom\bigl(\e^{-tH(A,V_{R})}\bigr) = \mathrm{L}^{2}(\mathbb{R}^{d})$
  such that 
  \begin{equation}
    \label{graphlimit}
    \lim_{R\to\infty} \bigl( \Vert \psi_{R} -\psi\Vert_{2} + \Vert
    \e^{-tH(A,V_{R})}\psi_{R} -\phi\Vert_{2} \bigr) = 0\,.
  \end{equation}
  According to Lemma~\ref{strongcvg} the convergence in
  (\ref{graphlimit}) holds for every
  $\psi\in\mathrm{L}^{2}_{\mathrm{G}}(\mathbb{R}^{d})$, if we set
  $\psi_{R} = \psi$ and $\phi=K_{t}\psi$, that is, 
  \begin{equation}
    \mathcal{G}_{t} \supseteq \bigl\{ (\psi,\phi)\in
    \mathrm{L}^{2}(\mathbb{R}^{d}) \times
    \mathrm{L}^{2}(\mathbb{R}^{d}) : \psi\in
    \mathrm{L}^{2}_{\mathrm{G}}(\mathbb{R}^{d}), \phi=K_{t}\psi\bigr\}\,.
  \end{equation}
  This implies $\mathrm{L}^{2}_{\mathrm{G}}(\mathbb{R}^{d}) \subseteq 
  \dom\bigl(\e^{-tH(A,V)}\bigr)$ and \eqref{fkieps}. Moreover, the 
  restriction of the self-adjoint operator $\e^{-tH(A,V)}$ to 
  $\mathrm{L}^{2}_{\mathrm{G}}(\mathbb{R}^{d})$ yields a symmetric operator. 
\end{proof}

Having settled Lemma~\ref{halfway}, we are in a position to
establish Theorem~\ref{fkigroup} on the semigroup properties of the
family $\{\e^{-tH(A,V)}\}_{t\ge 0}$.

\begin{proof}[of Theorem~\ref{fkigroup}]
  \begin{nummer}
  \item     
    The validity of the
    semigroup law (\ref{semigroupprop})
    on $\mathrm{L}^2_{\mathrm{G}}(\mathbb{R}^d)$ relies on the
    functional calculus for unbounded functions of unbounded
    self-adjoint operators,
    see e.g.\ Chap.~5 in \cite{BlEx94}, on Lemma~\ref{halfway} and on
    the inclusion $K_{t}\mathrm{L}^2_{\mathrm{G}}(\mathbb{R}^d)
    \subseteq \mathrm{L}^\infty_{\mathrm{G}}(\mathbb{R}^d)$, which was
    proven in Lemma~\ref{kate}. The latter two ensure that 
    both sides of (\ref{semigroupprop}) are well defined on
    $\mathrm{L}^2_{\mathrm{G}}(\mathbb{R}^d)$. 
  \item 
    Strong continuity of the
    orbit mapping $u_{\psi}$ for
    $\psi\in\mathrm{L}^2_{\mathrm{G}}(\mathbb{R}^d)$ follows 
    from the functional calculus, too, in that 
    \begin{equation}
      \label{strongcont}
      \Vert u_{\psi}(t+h) -u_{\psi}(t) \Vert_{2}^{2} =
      \int_{\mathbb{R}}\!\langle\psi,P(\d E)\psi\rangle \;
      (\e^{-(t+h)E} -\e^{-t E})^{2}
    \end{equation}
    for all $t\in[0,\infty[$ and all $h\in [-t,\infty[$. 
    Here $P$ denotes the projection-valued
    spectral measure of the Schr\"odinger operator $H:=H(A,V)$, that
    is, $P(I):=\Chi_{I}(H)$ for Borel sets 
    $I\subseteq\mathbb{R}$. Indeed, the integral in (\ref{strongcont})
    vanishes in the limit $h\to 0$ by the dominated-convergence
    theorem, because 
    we may assume $h\in [-t,h_0]$ with some $h_0 \in ]0,\infty[$ so that 
    the function $\mathbb{R} \ni E\mapsto (1+2\,\e^{-(t+h_0) E})^2$ 
    dominates the integrand of \eqref{strongcont} and is  
    $\langle\psi,P(\cdot)\psi\rangle$-integrable 
    due to $\psi\in \mathrm{L}^2_{\mathrm{G}}(\mathbb{R}^{d})$. 
    In the special case $t=0$, 
    this procedure gives the only meaningful right-sided limit 
    $h\downarrow 0$.
  \item 
    First we claim $\mathcal{C}_{0}^{\infty}(\mathbb{R}^{d}) \subset
    \dom(H \e^{-tH})$. Since $\mathcal{C}_{0}^{\infty}(\mathbb{R}^{d}) \subset
    \dom(\e^{-tH})$, this follows from Thm.\ 5.2.9(c) in
    \cite{BlEx94}, if 
    \begin{equation}
      \label{blankdomain}
      \int_{\mathbb{R}}\!\langle\varphi,P(\d E)\varphi\rangle \;
    (E\e^{-tE})^{2} < \infty
    \end{equation}
    for all $\varphi\in\mathcal{C}_{0}^{\infty}(\mathbb{R}^{d})$. The
    latter holds true, because $(E\e^{-tE})^{2} \le E^{2} +
    \e^{-2t_{0}E}$ for all $E\in\mathbb{R}$ with some $t_{0}>t$ and
    because $\mathcal{C}_{0}^{\infty}(\mathbb{R}^{d}) \subset
    \dom(H) \cap \dom(\e^{-t_{0}H})$. Next we compute the strong
    derivative of $u_{\varphi}$ for
    $\varphi\in\mathcal{C}_{0}^{\infty}(\mathbb{R}^{d})$. To this end,
    we consider the squared norm
    \begin{eqnarray}
      \lefteqn{\bigl\Vert h^{-1}\bigl(\e^{-(t +h) H}\varphi -
        \e^{-tH}\varphi\bigr)  + H\e^{-tH}\varphi \bigr\Vert_{2}^{2}}
      \nonumber  \\ 
      && \hspace*{1.5cm} = \int_{\mathbb{R}}\!\langle\varphi,P(\d
      E)\varphi\rangle \;  \bigl[h^{-1}\bigl(\e^{-(t+h) E}
      -\e^{-tE}\bigr) + E\e^{-tE} \bigr]^{2} \qquad \qquad     
      \label{squarednorm}
    \end{eqnarray}
    for $h \in ]-t,1] \setminus \{0\}$ and claim that it vanishes in the
    limit 
    $h \to 0$. (In the special case $t=0$, the limit gives the
    only meaningful right-sided derivative.) This follows from the
    dominated-convergence theorem and the $h$-independent upper
    bound $2E^{2}\bigl(2+ \e^{-2tE} + 2\e^{-2(t+1)E}\bigr)$ for the
    integrand in (\ref{squarednorm}). This bound is $\langle\varphi,
    P(\cdot) \varphi\rangle$-integrable as a function of $E$ because of
    $\varphi\in\mathcal{C}_{0}^{\infty}(\mathbb{R}^{d}) \subset \dom(H)$ and
    (\ref{blankdomain}). 
    
    It remains to show that $u_{\varphi}$ is the \emph{unique}
    solution of the initial-value problem (\ref{initval}). To this end,
    let $\Phi$ be an arbitrary solution of (\ref{initval}) and fix
    $t>0$ arbitrary. By the above reasoning one has 
    $\frac{\d}{\d s}\, \e^{-(t-s)H}g = H
    \e^{-(t-s)H}g$ in the strong sense for
    arbitrary $s\in]0,t[$ and arbitrary $g \in
    \mathcal{C}_{0}^{\infty}(\mathbb{R}^{d})$.  As a consequence, one
    finds
    \begin{equation}
      \frac{\d}{\d s}\, \langle \e^{-(t-s)H}g, \Phi(s)\rangle =
      \langle H\e^{-(t-s)H}g, \Phi(s)\rangle - \langle
      \e^{-(t-s)H}g, H \Phi(s)\rangle  =0\qquad
    \end{equation}
    by the assumptions on $\Phi$ and the self-adjointness of
    $H$. Hence, the fundamental theorem of calculus implies
    \begin{eqnarray}
      0 & = &  \int_{0}^{t}\!\d s\; \frac{\d}{\d s}\, \langle
      \e^{-(t-s)H}g, \Phi(s)\rangle = 
      \langle g, \Phi(t)\rangle - \langle \e^{-tH}g,
      \Phi(0)\rangle  \nonumber\\
      &= & \langle g, \Phi(t)\rangle - \langle g,
      \e^{-tH}\varphi\rangle = \langle g, \Phi(t) -u_{\varphi}(t)
      \rangle\,. 
    \end{eqnarray}
    The denseness of $\mathcal{C}_{0}^{\infty}(\mathbb{R}^{d})$ in 
    $\mathrm{L}^{2}(\mathbb{R}^{d})$ completes the proof of
    uniqueness. \qed
  \end{nummer}
\end{proof}

An immediate consequence of the just-proven Theorem~\ref{fkigroup} is

\begin{corollary}
  \label{core}
  Assume the situation of Theorem~\ref{fki}. Then
  $\mathrm{L}^2_{\mathrm{G}}(\mathbb{R}^{d})$ is an operator core for
  $\e^{-tH(A,V)}$ for all $t>0$.
\end{corollary}

\begin{proof}
  By Theorem~\ref{fkigroup} and the symmetry of $\e^{-tH(A,V)}$ on
  $\mathrm{L}^2_{\mathrm{G}}(\mathbb{R}^{d})$, see Lemma~\ref{halfway},
  all three assumptions of Thm.~1 
  in \cite{Nus70} are fulfilled by choosing there $\alpha =t \in\, ]0,\infty[$,
  $S_t = \e^{-t H(A,V)}$ with $\dom(S_t) = \mathrm{L}^2_{\mathrm{G}}
  (\mathbb{R}^{d})$ and $D=\mathrm{L}^2_{\mathrm{G}}(\mathbb{R}^{d})$.
  In this context, we recall from Lemma~\ref{halfway} that $\e^{-t H(A,V)}$
  is symmetric on $\mathrm{L}^2_{\mathrm{G}}(\mathbb{R}^{d})$ and from 
  Theorem~\ref{fkigroup}
  that the mapping $[0,\infty[ \ni t \mapsto \langle\psi,
  u_{\psi}(t)\rangle$ is continuous -- and hence Borel measurable --
  for every $\psi\in\mathrm{L}^2_{\mathrm{G}}(\mathbb{R}^{d})$ due to the
  strong continuity of the orbit mapping $u_{\psi}$. Therefore 
  the claim follows from Thm.~1 in \cite{Nus70}.
\end{proof}

The remaining part of the proof of Theorem~\ref{fki} is provided by

\begin{lemma}
  \label{carleman}
  Assume the situation of Theorem~\ref{fki} and let $K_{t}$ be defined
  as in Lemma~\ref{kate}. Then one has the equality
  \begin{equation}
    K_{t} = \e^{-t H(A,V)}\,.
  \end{equation}
\end{lemma}

\begin{proof}
  We follow \cite{AcGl81} or \cite{Sto70} and introduce the restriction 
  $K_{t}^{0}:=K_{t}|_{\dom(K_{t}^{0})}$ of the maximal Carleman operator
  $K_{t}$ to the subspace
  \begin{equation}
    \dom(K_{t}^{0}) := \bigl\{\psi\in\dom(K_{t}) : \kappa_{t} \psi\in
    \mathrm{L}^{1}(\mathbb{R}^{d})\bigr\}\,,
  \end{equation}
  where the function $\mathbb{R}^{d} \ni x \mapsto \kappa_{t}(x) := \Vert
  k_{t}(x,\cdot) \Vert_{2} = [ k_{2t}(x,x)]^{1/2}$ is well defined 
  and continuous because of Lemma~\ref{kCarleman}.
  The estimate (\ref{kbound}) in Lemma~\ref{bridgelemma} and 
  Remark \ref{lepsincl} imply $\mathrm{L}^{2}_{\mathrm{G}}(\mathbb{R}^{d})
  \subseteq \dom(K_{t}^{0})$. Thus, the Feynman-Kac-It\^{o}
  formula from Lemma~\ref{halfway} leads to
  \begin{equation}
    \label{carlstart}
    \e^{-tH(A,V)}|_{\mathrm{L}^{2}_{\mathrm{G}}(\mathbb{R}^{d})} 
    = K_{t}|_{\mathrm{L}^{2}_{\mathrm{G}}(\mathbb{R}^{d})}  
    = K_{t}^{0}|_{\mathrm{L}^{2}_{\mathrm{G}}(\mathbb{R}^{d})}
    \subseteq K_{t}^{0} \,.
  \end{equation}
  Here, as usual, the notation $A\subseteq B$ means
  that the operator $B$ is an extension of the operator $A$. By
  Thm.~10.1 in \cite{Sto70} the operator 
  $K_{t}^{0}$ is symmetric, hence closable. Taking the closure of
  (\ref{carlstart}) with respect to the graph norm and exploiting 
  Corollary~\ref{core}, we get
  $\e^{-tH(A,V)} \subseteq \overline{K_{t}^{0}}$.
  Since $K_{t}^{0}$ is symmetric, so is its closure 
  $\overline{K_{t}^{0}}$. Therefore we conclude 
  \begin{equation}
    \label{carl}
    \e^{-t H(A,V)} = \overline{K_{t}^{0}}\,,
  \end{equation}
  because self-adjoint operators are maximally symmetric. Furthermore,
  we observe the equalities $\overline{K_{t}^{0}} =
  \bigl(~\overline{K_{t}^{0}}~\bigr)^{*} = (K_{t}^{0})^{*} = K_{t}$, 
  which hold according to (\ref{carl}), Thm.~VIII.1(c) in \cite{ReSi80} and
  Thm.~10.1 in \cite{Sto70}. This completes the proof. 
\end{proof}

Finally, we gather our previous results to complete the 

\begin{proof}[of Theorem~\ref{fki}]
  Corollary~\ref{core} has
  established that $\mathrm{L}^2_{\mathrm{G}}(\mathbb{R}^{d})$ is an
  operator core for $\e^{-tH(A,V)}$. The remaining assertions of
  Theorem~\ref{fki} follow from Lemma~\ref{carleman},
  Lemma~\ref{kate} and Lemma~\ref{kCarleman}.
\end{proof}


%
%
\section{Proofs of Theorem \protect{\lowercase{\ref{intkernel}}}, 
  Corollary \protect{\lowercase{\ref{speckernel}}} and 
  Corollary \protect{\lowercase{\ref{kernelcalculus}}}}
\label{intkernelproof}

The following lemma is in the spirit of Thm.~B.7.8 in \cite{Sim82},
but, among others, we do not assume that the operator $M$ is bounded. 
 
\begin{lemma}
  \label{TBT}
  Let $M$ be the maximal self-adjoint Carleman operator induced by the 
  Borel-measurable and Hermitian integral kernel $m: \mathbb{R}^d \times
  \mathbb{R}^d \rightarrow \mathbb{C}$ in the sense that 
  \begin{eqnarray}
    \mathcal{C}_0^\infty(\mathbb{R}^d) & \subset &  \dom(M)  := 
    \Bigl\{\psi\in\mathrm{L}^{2}(\mathbb{R}^{d}):
      \int_{\mathbb{R}^{d}}\d{}y\; m(\cdot,y)\,\psi(y) \in
      \mathrm{L}^{2}(\mathbb{R}^{d}) \Bigr\}   \,,
    \nonumber\\
    M\psi & = & \int_{\mathbb{R}^{d}}\!\d{}y\; m(\cdot,y)\,
    \psi(y)
    \label{Tkernel}
  \end{eqnarray}
  for all $\psi \in \dom(M)$,  
  $m(x,y)=m^*(y,x)$ for Lebesgue-almost all pairs $(x,y) \in
  \mathbb{R}^{d} \times \mathbb{R}^{d}$ and $m$ has the Carleman property
  \eqref{carledef}. 
  Assume further that $x \mapsto m(\cdot, x)$ defines a strongly continuous
  mapping from $\mathbb{R}^{d}$ to $\mathrm{L}^{2}(\mathbb{R}^{d})$. 
  Finally, let $B$ be a bounded operator on
  $\mathrm{L}^{2}(\mathbb{R}^{d})$  such that $M\!B$ and $M\!B^*$ 
  are also bounded and that $M\!BM$ admits a bounded closed 
  extension $\overline{M\!BM}$ to all of $\mathrm{L}^{2}(\mathbb{R}^{d})$.
  Then
  \begin{nummer}
  \item \label{intlemmaone}
    $\overline{M\!BM}$ is a bounded Carleman operator induced by the
    continuous integral kernel 
    $\beta:\mathbb{R}^{d}\times\mathbb{R}^{d} \to\mathbb{C}$,  
    $(x,y) \mapsto \beta(x,y) := \langle m(\cdot,x), B
    m(\cdot,y)\rangle$ in the sense that 
    \begin{equation}
      \label{TBTkernel}
      \overline{M\!BM}\psi = \int_{\mathbb{R}^{d}}\!\d{}y\; \beta(\cdot,y)
      \, \psi(y)  
    \end{equation}
    for all $\psi \in \mathrm{L}^{2}(\mathbb{R}^{d})$ and 
    $\beta$ has the Carleman property \eqref{carledef}. 
  \item \label{intlemmatwo}
    the left-hand side of \eqref{TBTkernel} has a continuous representative 
    in $\mathrm{L}^{2}(\mathbb{R}^{d})$, which is given by the
    right-hand side of \eqref{TBTkernel}. 
  \item \label{intlemmathree}
    for any $w\in\mathrm{L}^{\infty}(\mathbb{R}^{d})$ with 
    $\int_{\mathbb{R}^d\times\mathbb{R}^d}\d x\d y\, |w(x)|^2\,
    |m(x,y)|^2 < \infty$
    the product $\overline{M\!BM}\hat{w}$ is a Hilbert-Schmidt
    operator with squared norm given by
    \begin{equation}
      \label{TBTtrace}
      \mathrm{Trace}\bigl[ \hat{w}^{*} | \overline{M\!BM} |^{2}
      \hat{w} \bigr] = 
      \int_{\mathbb{R}^{d}}\!\d{}x\; |w(x)|^{2} 
      \int_{\mathbb{R}^{d}}\!\d{}y\;|\beta(x,y)|^{2}\,. \quad
    \end{equation}
    Here $\hat{w}$ is the bounded multiplication operator uniquely
    corresponding to $w$, and $\hat{w}^{*}$ denotes its Hilbert adjoint. 
  \end{nummer}
\end{lemma}

\begin{proof}
  The strong continuity of the mapping 
  $\mathbb{R}^{d}\to\mathrm{L}^{2}(\mathbb{R}^{d})$, $x \mapsto m(\cdot, x)$,
  the triangle and the Cauchy-Schwarz inequality imply the
  continuity of the function $\mathcal{M}:\mathbb{R}^{d}\to \mathbb{R}$, 
  $x \mapsto \mathcal{M}(x):= \| m(\cdot, x)\|_2$ because 
  $|\mathcal{M}(x) - \mathcal{M}(x')| \le \|m(\cdot,x) -
  m(\cdot,x')\|_2$. 
  Now, for every $\varphi\in \mathcal{C}_0^\infty(\mathbb{R}^{d})$ and every
  $\psi\in\mathrm{L}^{2}(\mathbb{R}^{d})$
  the Cauchy-Schwarz inequality provides the estimate
  \begin{equation}
    \int_{\mathbb{R}^{d}\times\mathbb{R}^{d}}\!\!\
      \d x\d{}y\; |\psi(y)|\, |m(y,x)|\, |\varphi(x)|
    \leq \|\psi\|_2\, \|\varphi\|_2 \,
    \|\mathcal{M}\Chi_{\mathrm{supp}\,\varphi}\|_2  < \infty\qquad
  \end{equation}
  due to the continuity of $\mathcal{M}$. Therefore, \eqref{Tkernel} and
  Fubini's theorem yield
  \begin{equation}
    \label{Tstart}
    \langle M\varphi,\psi\rangle = \int_{\mathbb{R}^{d}}\!\d{}x\; 
    \varphi^{*}(x)\,\langle m(\cdot,x),\psi\rangle\,,
  \end{equation}
  where the scalar product in the integrand is well defined,
  because, by hypothesis, $m(\cdot,
  x)\in\mathrm{L}^{2}(\mathbb{R}^{d})$ for all $x\in\mathbb{R}^{d}$. 
  Next, we consider a sequence $(\psi_{n})_{n\in\mathbb{N}} \subset
  \mathcal{C}_{0}^{\infty}(\mathbb{R}^{d})$ with $\lim_{n\to\infty} \|
  \psi_{n} - \psi\|_{2} =0$ and $\sup_{n\in\mathbb{N}}\{\|\psi_{n}\|_{2}\}
  \le 2\|\psi\|_{2}$. From the boundedness of $\overline{M\!BM}$, the
  continuity of the scalar product $\langle\cdot,\cdot\rangle$ and
  \eqref{Tstart} we conclude  
  \begin{eqnarray}
    \langle\varphi, \overline{M\!BM}\psi\rangle & = & 
    \lim_{n\to\infty} \langle\varphi, {M\!BM}\psi_{n}\rangle\nonumber\\*
    & = & \lim_{n\to\infty} \langle M\varphi, BM\psi_{n}\rangle\nonumber\\*
    & = & \lim_{n\to\infty} \int_{\mathbb{R}^{d}}\!\d{}x\; \varphi^{*}(x)\, 
    \langle m(\cdot,x), BM\psi_{n}\rangle \nonumber\\*
    & = & \lim_{n\to\infty} \int_{\mathbb{R}^{d}}\!\d{}x\; \varphi^{*}(x)\, 
    \langle M\!B^* m(\cdot,x), \psi_{n}\rangle \,.
    \label{firstdense}
  \end{eqnarray}
  Since 
  \begin{equation}
    \label{bracketbound}
    \sup_{n\in\mathbb{N}} \bigl|\langle M\!B^* m(\cdot,x),
    \psi_{n}\rangle \bigr|
    \le 2\,\|M\!B^{*}\| \, \|\psi\|_{2} \, \mathcal{M}(x)    
  \end{equation}
  for all $x\in\mathbb{R}^{d}$, $M\!B^{*}$ is bounded and $\mathcal{M}$
  is continuous, the dominated-convergence theorem and the continuity
  of the scalar product yield
  \begin{equation}
    \langle\varphi, \overline{M\!BM}\psi\rangle =
    \int_{\mathbb{R}^{d}}\!\d{}x\; \varphi^{*}(x)\,  
    \langle M\!B^* m(\cdot,x), \psi\rangle
  \end{equation}
  for all $\varphi\in\mathcal{C}_0^\infty(\mathbb{R}^{d})$ and all
  $\psi\in\mathrm{L}^{2}(\mathbb{R}^{d})$. Moreover, the function
  $\mathbb{R}^{d}\ni x\mapsto \langle M\!B^* m(\cdot,x), \psi\rangle$ 
  belongs to $\mathrm{L}^{\infty}_{\mathrm{loc}}(\mathbb{R}^{d})$,
  confer \eqref{bracketbound}, so that the lemma of Du 
  Bois-Reymond -- also known as the fundamental lemma 
  of the calculus of  variations, see e.g.\ Lemma~3.26 in
  \cite{Ada75} -- implies
  \begin{eqnarray}
    \bigl( \overline{M\!BM}\psi\bigr)(x) & = &
    \langle M\!B^* m(\cdot,x), \psi\rangle \nonumber\\*
    & = & \int_{\mathbb{R}^{d}}\!\d{}y \left[
      \int_{\mathbb{R}^{d}}\!\d{}z\; m(y,z) \, \bigl(
      B^{*}m(\cdot,x)\bigr)(z)  \right]^{*}\; \psi(y) \nonumber\\*
    & = & \int_{\mathbb{R}^{d}}\!\d{}y \; \langle m(\cdot,x), B
    m(\cdot,y) \rangle \, \psi(y)
    \label{TBTlimit}
  \end{eqnarray}
  for Lebesgue-almost all $x\in\mathbb{R}^{d}$ and all
  $\psi\in\mathrm{L}^{2}(\mathbb{R}^{d})$. To get the last equality,
  we have also used the Hermiticity, $m(x,y)=m^{*}(y,x)$
  for Lebesgue-almost all pairs
  $(x,y)\in\mathbb{R}^{d}\times\mathbb{R}^{d}$. This proves
  \eqref{TBTkernel}. 

  The Carleman property \eqref{carledef} for $\beta$  
  follows from part
  \itemref{intlemmathree} of 
  the lemma (to be proven below). Indeed, since $m$ is Hermitian
  and since $\mathcal{M}$ is continuous, one may choose $w=\Chi_{\Lambda}$ in
  \eqref{TBTtrace} 
  for an arbitrary bounded Borel subset
  $\Lambda\subset\mathbb{R}^{d}$. This completes the proof of 
  part~\itemref{intlemmaone}.

  The proof of assertion~\itemref{intlemmatwo} follows from the first
  equality in \eqref{TBTlimit},
  the fact that the mapping $\mathbb{R}^{d}\to\mathrm{L}^{2}(\mathbb{R}^{d})$, 
  $x \mapsto m(\cdot, x)$, is strongly continuous, $M\!B^*$ is bounded and
  $\langle\cdot,\cdot\rangle$ is continuous.

  For the proof of assertion \itemref{intlemmathree} we exploit our
  assumption on $w$, the maximality of the Carleman operator $M$,
  \eqref{Tkernel} and Thm.\ VI.23 in \cite{ReSi80} to conclude 
  that $M\hat{w}$ is Hilbert-Schmidt. Therefore, $M\!BM\hat{w} =
  \overline{M\!BM}\hat{w}$ 
  is Hilbert-Schmidt, too, by the boundedness of $M\!B$ and the 
  H{\"o}lder inequality for Schatten norms, see e.g.\ Thm.~2.8 in 
  \cite{Sim79b}. Thanks to $w\in\mathrm{L}^{\infty}(\mathbb{R}^{d})$
  and Eq.~\eqref{TBTkernel} we have 
  $\overline{M\!BM}\hat{w}\psi = \int_{\mathbb{R}^d}  
  \d y\, \beta(\cdot,y)\, w(y)\,\psi(y)$ for all 
  $\psi\in\mathrm{L}^{2}(\mathbb{R}^{d})$. Hence \eqref{TBTtrace}
  follows from an anew application of Thm.\ VI.23 in \cite{ReSi80}.
\end{proof}

After these preparations it is easy to deduce Theorem~\ref{intkernel} as 
a special case.

\begin{proof}[of Theorem \ref{intkernel}]
  We apply Lemma~\ref{TBT} with the choices 
  $M=\e^{-t H(A,V)}$ and $B=\e^{{2t} H(A,V)}F\bigl(H(A,V)\bigr)$, where
  $t\in]0,\tau/2[$. 

  This is allowed, because Theorem~\ref{fki} ensures  
  that $\e^{-tH(A,V)}$ is a maximal Carleman operator with the
  required properties, recall Remark~\ref{lepsdense},
  Lemma~\ref{bridgelemma} and Remark~\ref{Carleswap}. 

  Furthermore, we observe from \eqref{Fcond} and the
  functional calculus for unbounded functions of unbounded
  self-adjoint operators, see e.g.\ Chap.~5 in \cite{BlEx94}, that the
  operator product 
  $B=\e^{{2t} H(A,V)}F\bigl(H(A,V)\bigr)$ is bounded. The functional
  calculus also guarantees that the 
  two operator products $M\!B$ and $M\!B^*$ are bounded
  and that the equality $M\!BM = F\bigl(H(A,V)\bigr)$  holds on 
  $\dom(M)$. The latter implies the boundedness of
  $\overline{M\!BM}=F\bigl(H(A,V)\bigr)$, because
  $F\in\mathrm{L}^\infty(\mathbb{R})$. 

  Finally, the finiteness of the integral $\int_{\mathbb{R}^{d}\times 
    \mathbb{R}^{d}} \d{}x\d{}y\; |w(x)|^2 |k_{t}(x,y)|^{2}$
  for all $w\in\mathrm{L}^\infty_\mathrm{G}(\mathbb{R}^d)$ follows from the
  estimate \eqref{kbound} with sufficiently small $\delta >0$, 
  inequality \eqref{elineq} and Remark~\ref{lepsincl}. 
  Thus, all assumptions of Lemma~\ref{TBT}
  are fulfilled and Theorem~\ref{intkernel} holds with $f=\beta$ and
  for all $w\in\mathrm{L}^\infty_\mathrm{G}(\mathbb{R}^d)$.
\end{proof}

Next we show how to deduce
Corollary~\ref{speckernel} from Theorem~\ref{intkernel}.

\begin{proof}[of Corollary~\ref{speckernel}]
  Clearly, choosing $F=\Chi_{I}$ in Theorem~\ref{intkernel} is in
  accordance with \eqref{Fcond} because of $\sup I <
  \infty$. Therefore, part~\itemref{intkernelexist} of this 
  theorem  yields the existence and continuity of the integral kernel
  $p_{I}$ of $\Chi_{I}\bigl(H(A,V)\bigr)$. To  derive \eqref{tracechi} 
  we note that the operator
  $\hat{w}^{*}\Chi_{I}\bigl(H(A,V)\bigr)\hat{w}$ is trace 
  class by Theorem~\ref{intkernelHStrace} and $\Chi_{I}^{2}=\Chi_{I}$.  
  Moreover, thanks to $w\in\mathrm{L}^{\infty}_{\mathrm{G}}(\mathbb{R}^{d})$ 
  the $\mathrm{L}^{2}(\mathbb{R}^{d}\times\mathbb{R}^{d})$-function
  $(x,y) \mapsto w^{*}(x)p_{I}(x,y)w(y)$ is an integral kernel for
  $\hat{w}^{*}\Chi_{I}\bigl(H(A,V)\bigr)\hat{w}$. 
  Recalling that $\Lambda_{\ell}(x)$ is the open cube in $\mathbb{R}^{d}$ with
  edge length $\ell >0$ and centre $x\in\mathbb{R}^{d}$, an
  application of Thm.~3.1 in \cite{Bri88}, see also \cite{Bri90} or
  \cite{Bir89}, gives the equality
  \begin{eqnarray}
    \lefteqn{ \mathrm{Trace}\bigl[
      \hat{w}^{*}\Chi_{I}\bigl(H(A,V)\bigr) \hat{w} \bigr] } \nonumber\\*
    && =  \int_{\mathbb{R}^{d}}\d{}x\; \lim_{\ell\downarrow 0} \,
    \ell^{-2d} \int_{\Lambda_{\ell}(x) \times \Lambda_{\ell}(x)} \!
    \d{}x'\d{}y'\;
     w^{*}(x')\, p_{I}(x',y') \, w(y') \,. \qquad
    \label{brislawneq}
  \end{eqnarray}
  The continuity of $p_{I}$ and the Lebesgue
  differentiation theorem, see e.g.\ Sects.~I.1.3 and I.1.8 in
  \cite{Ste70}, now complete the proof because
  \begin{eqnarray}
    \lefteqn{\lim_{\ell\downarrow 0} \,
    \ell^{-2d} \int_{\Lambda_{\ell}(x) \times \Lambda_{\ell}(x)} \!
    \d{}x'\d{}y'\;
     w^{*}(x')\, p_{I}(x',y') \, w(y')} \nonumber\\* 
    && \hspace*{2cm} = p_{I}(x,x) \; \lim_{\ell\downarrow 0} \biggl| \ell^{-d}
    \int_{\Lambda_{\ell}(x)} \!\d{}x'\; w(x') \biggr|^{2}
    \nonumber\\*
    && \hspace*{2cm} = p_{I}(x,x) \, |w(x)|^{2}
  \end{eqnarray}
  for Lebesgue-almost all $x\in\mathbb{R}^{d}$.
\end{proof}

Now we are concerned with the second corollary to Theorem~\ref{intkernel}.

\begin{proof}[of Corollary~\ref{kernelcalculus}]
  We fix $x,y\in \mathbb{R}^d$. In the first case
  we apply the functional calculus 
  to the right-hand side of \eqref{fexeq}. This gives
  \begin{equation}
    \label{ffuca}
    f(x,y) = \int_{\mathbb{R}}\!\d\vartheta_t(E;x,y)\; \e^{2tE}\,F(E)
  \end{equation}
  for any $t \in  ]0,\tau/2[$ with the complex spectral ``distribution''
  function 
  $\vartheta_t(E;x,y) := \bigl\langle k_t(\cdot,x), 
  \Chi_{]-\infty,E[}\bigl(H(A,V)\bigr) k_t(\cdot,y)\bigr\rangle$.
  Here, $\tau >0$ is the constant required to exist for $F$ in \eqref{Fcond}.
  In particular, for $F=\Chi_{]-\infty,E_0[}$ with $E_0\in\mathbb{R}\,$, Eq.\
  \eqref{ffuca} takes the form
  \begin{equation}
    \label{absstetig}
    p(E_0;x,y) = \int_{-\infty}^{E_0}\!\d\vartheta_t(E;x,y)\; \e^{2tE}\,.
  \end{equation}
  This equation holds for arbitrary $t>0$, because $\tau$ can be chosen
  arbitrarily large in this particular case.
  Taken together, \eqref{ffuca} and \eqref{absstetig}
  yield the claim \eqref{fucakernel}.

  In the second case we may write 
  \begin{equation}
    k_t(x,y) = \bigl\langle k_{t/2}(\cdot,x), k_{t/2}(\cdot,y)\bigr\rangle
    = \int_{\mathbb{R}}\!\d\vartheta_{t/2}(E;x,y)
    = \int_{\mathbb{R}}\!\d p(E;x,y)\; \e^{-tE}
  \end{equation}
  for all $t>0$.
  Here, the first equality is due to the Hermiticity and the semigroup 
  property of the kernel $k_t$, the second equality is just the definition 
  of $\vartheta_{t/2}$ and the last equality follows from \eqref{absstetig}.
\end{proof}

For convenience, we formulate and prove simple estimates on the integral
kernel of a spectral projection in the remainder of this section. We will only
need these estimates for the applications to random
Schr\"odinger operators.

\begin{lemma}
  \label{kernelineq}
  Assume the situation of Corollary~\ref{speckernel}.
  Then the diagonal of the continuous integral kernel
  $p_I$ of the spectral projection $\Chi_I\bigl(H(A,V)\bigr)$ obeys the 
  estimates
  \begin{equation}
    \label{kernelest}
    0 \le p_I(x,x) \le \e^{t \sup I} \, k_t(x,x)
  \end{equation}
  for all $x \in\mathbb{R}^d$ with any $t \in]0,\infty[$.
\end{lemma}

\begin{proof}
  Fix $x\in\mathbb{R}^{d}$ arbitrary, pick $\varphi \in
  \mathcal{C}_{0}^{\infty}(\mathbb{R}^{d})$ and define
  $\varphi_{x}^{(\varepsilon)}$ by $\varphi_{x}^{(\varepsilon)}(y) :=
  \varepsilon^{-d} \varphi\bigl((y-x)/\varepsilon\bigr)$ for every
  $y\in\mathbb{R}^{d}$ and every $\varepsilon\in ]0,1]$. Then
  $\{\varphi_{x}^{(\varepsilon)}\}_{\varepsilon\in ]0,1]} \subset
  \mathrm{L}^{2}(\mathbb{R}^{d})$ is a family of approximating delta functions
  at $x\in\mathbb{R}^{d}$. By the continuity of 
  $p_I$ and the dominated-convergence theorem one gets the representation
   \begin{equation}
     p_I(x,x) = \lim_{\varepsilon\downarrow 0}
     \langle \varphi_{x}^{(\varepsilon)}, \Chi_I\bigl(H(A,V)\bigr)
     \varphi_{x}^{(\varepsilon)} \rangle \,.
   \end{equation}
   The same arguments yield
   \begin{equation}
     k_t(x,x) = \lim_{\varepsilon\downarrow 0}
     \langle \varphi_{x}^{(\varepsilon)}, \e^{-t H(A,V)}
     \varphi_{x}^{(\varepsilon)} \rangle 
   \end{equation}
   for any $t \in ]0,\infty[$.  The claim \eqref{kernelest} now follows from
   the functional calculus and the elementary inequalities
   \begin{equation}
     0 \le \Chi_I(E) \le \e^{t (\sup I -E)}
   \end{equation}
   for all $E\in\mathbb{R}\,$.
\end{proof}


%
\section{Proofs of Lemma \protect{\lowercase{\ref{rpot}}}, Corollary
\protect{\lowercase{\ref{doscor}}} and Corollary
\protect{\lowercase{\ref{avsemi}}}} 
\label{rpotproof}

\begin{proof}[of Lemma \ref{rpot}]
  We mimic the proof of \cite{KiMa83}, see also Prop.~V.3.2 in \cite{CaLa90}. 
  By the definition of $p(d)$ in property \ass{S} and 
  since $(d/2)p_1/[p_1-p(d)] < p_{2}$, 
  we can find $\nu\in]0,2[$ and $r\in]p(d), p_1[$ such that 
  \begin{equation}
    \label{exponents}
    \frac{d}{\nu} \,\frac{p_1}{p_{1}-r} < p_{2}\,.
  \end{equation}
  Next, we pick a constant $c\in]0,\infty[$ and define 
  \begin{eqnarray}
    V_2^{(\omega)}(x) & := & V^{(\omega)}(x) \, 
    \Theta\bigl( c(1+|x|^{\nu}) -
    |V^{(\omega)}(x)|\bigr)\,, \mathletter{a} \label{Wdef} \\*
    V_1^{(\omega)}(x) & := &  V^{(\omega)}(x) - V^{(\omega)}_2(x)
    \mathletter{b}
  \end{eqnarray}
  for all $\omega\in\Omega$ and all $x\in\mathbb{R}^d$.
  Clearly, for every $\omega\in\Omega$ the realization 
  $V_2^{(\omega)}$ satisfies \eqref{Wcond} for all $\varepsilon >0$. 
  We will show below that
  $V_1^{(\omega)}\in\mathrm{L}^{r}_{\mathrm{unif, loc}}(\mathbb{R}^{d})$ for 
  $\mathbb{P}$-almost all $\omega\in\Omega$.
  This proves the lemma, because
  $\mathrm{L}^{r}_{\mathrm{unif, loc}}(\mathbb{R}^{d}) \subseteq
  \mathcal{K}(\mathbb{R}^{d})$, see e.g.\ Eq.\ (A$\,$21) in \cite{Sim82}
  for $d\geq 2$ and note 
  $\mathcal{K}(\mathbb{R}) = \mathrm{L}^{1}_{\mathrm{unif, loc}}
  (\mathbb{R})$.   

  In this proof we use the abbreviation $\Lambda(y):= \Lambda_{1}(y)$
  for the open unit cube in $\mathbb{R}^{d}$ with centre
  $y\in\mathbb{R}^{d}$. 
  To prove $\mathbb{P}\bigl[ V_1\in \mathrm{L}^{r}_{\mathrm{unif,
      loc}}(\mathbb{R}^{d})\bigr] =1$ we apply the
  ``Chebyshev-Markov'' inequality $\Theta(\xi-1) \leq |\xi|^{\kappa}$ with
  $\kappa=p_{1}-r >0$ to obtain for all $\omega\in\Omega$ the estimate
  \begin{equation}
    \| V^{(\omega)}_1\Chi_{\Lambda(y)}\|_{r}^{r} 
    = \int_{\Lambda(y)}\d{}x\, |V^{(\omega)}(x)|^{r}\, \Theta\left(
      \frac{|V^{(\omega)}(x)|}{c (1+|x|^{\nu})}  -1 \right)
    \leq
    \frac{\tilde{c}^{r}\, \|
      V^{(\omega)}\Chi_{\Lambda(y)}\|_{p_1}^{p_1}}{(1+|y|^{\nu})^{p_{1}-r}} 
  \end{equation}
  for all $y\in\mathbb{Z}^{d}$ with some
  constant $\tilde{c} \in]0,\infty[$, which is independent of
  $y\in\mathbb{Z}^{d}$. 
  This implies
  \begin{eqnarray} 
    \sum_{y\in\mathbb{Z}^{d}} \mathbb{P}\bigl[\|
      V_1\Chi_{\Lambda(y)}\|_{r} > 1 \bigr] 
    &\leq& \sum_{y\in\mathbb{Z}^{d}} \mathbb{E}\left[ 
      \Theta\left(  \frac{\tilde{c}\;
          \| V\Chi_{\Lambda(y)}\|_{p_1}^{p_1/r}}%
        {(1+|y|^{\nu})^{(p_1 -r)/r}}  -1
      \right)\right] \nonumber\\*
    &\leq& \tilde{c}^{q}  \sum_{y\in\mathbb{Z}^{d}}  
      \frac{\mathbb{E}\left[ \| V\Chi_{\Lambda(y)}\|_{p_1}^{p_1q/r}\right]}%
      {(1 + |y|^{\nu})^{(p_{1}-r)q/r}}\,.
      \label{unifchain}
  \end{eqnarray}
  In order to get the second inequality in \eqref{unifchain}, we used
  the ``Chebyshev-Markov'' inequality with $\kappa=q$, 
  where $q$ is chosen such that 
  \begin{equation}
    \label{moreexponents}
    \frac{d}{\nu} \,
    \frac{p_1}{p_{1}-r} < \frac{p_1 q}{r} < p_2 \,.
  \end{equation}
  The numerator in the second line of \eqref{unifchain} is uniformly
  bounded in $y\in\mathbb{Z}^{d}$ due to the right inequality in 
  \eqref{moreexponents}, Jensen's inequality and property \ass{S}.
  The left inequality in \eqref{moreexponents} then assures that the 
  series in the second line of \eqref{unifchain} is summable, which
  implies by the first Borel-Cantelli lemma
  \begin{equation}
    \mathbb{P}\bigl[ \| V_1\Chi_{\Lambda(y)}\|_{r} > 1
    \mbox{~~for infinitely many~~} y\in\mathbb{Z}^{d} \bigr] =0\,.
  \end{equation}
  This delivers
  \begin{eqnarray}
    \mathbb{P}\left[ \sup_{ y\in\mathbb{Z}^{d}} \|
    V_1\Chi_{\Lambda(y)}\|_{r} = \infty \right]  & = &
    \mathbb{P}\bigl[ \| V_1\Chi_{\Lambda(y_0)}\|_{r} = \infty 
    \mbox{~~for some~~} y_0\in\mathbb{Z}^{d} \bigr] \nonumber\\*
    & \leq & \sum_{ y\in\mathbb{Z}^{d}} \mathbb{P}\bigl[ \|
    V_1\Chi_{\Lambda(y)}\|_{r} = \infty \bigr] \nonumber\\
    & \leq & \sum_{ y\in\mathbb{Z}^{d}} \mathbb{P}\bigl[ \|
    V\Chi_{\Lambda(y)}\|_{p_1} = \infty \bigr] \nonumber\\*
    & = & 0\,,
    \label{BoCa}
  \end{eqnarray}
  where we have used the countable subadditivity of $\mathbb{P}$ for the first 
  inequality and $|V_1| \leq |V|$ as well as $r<p_1$ for the second 
  inequality. The last equality in (\ref{BoCa}) follows from property
  \ass{S}. Thus, we have shown 
  \begin{equation}
    \mathbb{P}\bigl[ V_1\in \mathrm{L}^{r}_{\mathrm{unif,
        loc}}(\mathbb{R}^{d})\bigr] =1 \,. \qquad\qed
  \end{equation}
\end{proof}

For the proof of Corollary~\ref{doscor} and Corollary~\ref{avsemi}
we need suitable measurability properties of the involved integral kernels, 
which we establish in 

\begin{lemma}
  \label{messbar}
  Let $A$ be a vector potential with property \ass{A} and let $V$ be a 
  random scalar potential with property~\ass{S}. Then there exists 
  $\Omega_0 \in \mathcal{A}$ with $\mathbb{P}(\Omega_0)=1$ such that for
  every $\omega\in\Omega_0$
  \begin{nummer}
    \item
      \label{kmeas}
      the operator exponential $\e^{-t H(A,V^{(\omega)})}$ has a continuous 
      integral kernel $k_t^{(\omega)}$ for any $t>0$ and the mapping 
      \begin{equation}
        \begin{array}{ccc}
          \Omega_0 \times ]0,\infty[ \times \mathbb{R}^d \times \mathbb{R}^d
          & \rightarrow & \mathbb{C} \nonumber\\
          (\omega,t,x,y) & \mapsto & k_t^{(\omega)}(x,y)
        \end{array}
      \end{equation}
      is $\mathcal{A}_0 \otimes \mathcal{B}(]0,\infty[) \otimes
      \mathcal{B}(\mathbb{R}^d) \otimes \mathcal{B}(\mathbb{R}^d)$-measurable.
    \item 
      \label{pmeas}
      the spectral projection $\Chi_{]-\infty,E[}\bigl( 
      H(A,V^{(\omega)})\bigr)$ has a continuous integral kernel 
      $p^{(\omega)}(E;\cdot,\cdot)$ for any $E\in\mathbb{R}$ and the mapping 
      \begin{equation}
        \begin{array}{ccc}
          \Omega_0 \times \mathbb{R} \times \mathbb{R}^d \times \mathbb{R}^d
          & \rightarrow & \mathbb{C} \nonumber\\
          (\omega,E,x,y) & \mapsto & p^{(\omega)}(E;x,y)
        \end{array}
      \end{equation}
      is $\mathcal{A}_0 \otimes \mathcal{B}(\mathbb{R}) \otimes
      \mathcal{B}(\mathbb{R}^d) \otimes \mathcal{B}(\mathbb{R}^d)$-measurable.
    \end{nummer}
    Here, $\mathcal{A}_0$ is the restriction of the sigma-algebra 
    $\mathcal{A}$ of $\Omega$ to $\Omega_0$, and given any Borel set 
    $B\subseteq \mathbb{R}^d$ we denote by $\mathcal{B}(B)$ the 
    sub-sigma-algebra of Borel sets in $\mathbb{R}^d$ which are contained 
    in $B$. 
\end{lemma}

\begin{proof}
  The existence and continuity of the integral kernels is guaranteed by
  Corollary~\ref{rcor}, Lemma~\ref{bridgelemma}, Theorem~\ref{fki} and
  Corollary~\ref{speckernel} (see also Corollary~\ref{kernelcalculus}). The
  measurability claimed in \itemref{kmeas} follows from the Brownian-bridge
  representation \eqref{kern} for $k_t^{(\omega)}$. The claim of
  \itemref{pmeas} follows from \itemref{kmeas}, Corollary~\ref{kernelcalculus}
  and the invertibility of the Laplace transformation.
\end{proof}

\begin{proof}[of Corollary \ref{doscor}]
  We fix $E\in\mathbb{R}$ arbitrary. Lemma~\ref{pmeas} guarantees the
  existence, continuity and suitable measurability properties of the 
  integral kernel $p^{(\omega)}(E;\cdot,\cdot)$ of the spectral projection
  $\Chi_{]-\infty,E[}\bigl(H(A,V^{(\omega)})\bigr)$ for all
  $\omega\in\Omega_0\in\mathcal{A}$ with $\mathbb{P}(\Omega_{0})=1$. 
  Eq.\ (\ref{tracechi}) and
  Proposition~\ref{dosdef} imply that 
  \begin{equation}
    N(E) = \mathbb{E} \biggl[ \int_{\Gamma} \frac{\d
      x}{|\Gamma|} \; p(E;x,x) \biggr]
  \end{equation}
  is finite. Now the claim follows
  from Fubini's
  theorem, because $p^{(\omega)}(E;x,x) \ge 0$ for all
  $\omega\in\Omega_{0}$ and all $x\in\mathbb{R}^{d}$, see 
  Lemma~\ref{kernelineq}, and because
  $\mathbb{E}[p(E;x,x)]$ is independent of
  $x\in\mathbb{R}^{d}$ due to the $\mathbb{R}^{d}$-ergodicity of $V$.
\end{proof}

\begin{proof}[of Corollary \ref{avsemi}]
  We fix $t>0$ arbitrary. Lemma~\ref{kmeas} guarantees the
  existence, continuity and suitable measurability properties of the 
  integral kernel $k_t^{(\omega)}$ of the operator exponential
  $\e^{-t H(A,V^{(\omega)})}$ for all
  $\omega\in\Omega_0\in\mathcal{A}$ with $\mathbb{P}(\Omega_{0})=1$. 
  Jensen's inequality, Fubini's theorem and property \ass{L} imply for
  $\mu_{x,y}^{0,t}$-almost every path $b$ of the Brownian bridge
  the estimate 
  \begin{equation}
    \label{jenscons}
    \mathbb{E}\left[ \exp\biggl\{ - \int_{0}^{t}\!\d s\;
      V(b(s))\biggr\}\right]  \le
    \int_{0}^{t}\frac{\d s}{t}\; \mathbb{E}\bigl[ \exp\bigl\{ -t
      V(b(s))\bigr\}\bigr] \le \mathcal{L}_{t} < \infty
      \,, \qquad
  \end{equation}
  which shows that the integral kernel $\overline{k_{t}}$ is
  well defined and obeys the inequality   
  \begin{equation}
    \label{freekernelbound}
    |\overline{k_{t}}(x,y)| \le \mathbb{E} \bigl[
    |{k_{t}}(x,y)|\bigr] \le \mathcal{L}_{t} \; 
    \frac{\e^{-|x-y|^{2}/(2t)}}{(2\pi t)^{d/2}} 
  \end{equation}
  for all $x,y\in\mathbb{R}^{d}$, thereby proving \eqref{ueberfluessig}.
  The Hermiticity of $\overline{k_{t}}$ is inherited from that of
  $k_{t}$, see Lemma~\ref{kerneldef}. The estimate
  \eqref{freekernelbound} also yields
  $\overline{k_{t}}(x,\cdot) \in
  \mathrm{L}^{\infty}_{\mathrm{G}}(\mathbb{R}^{d})$ for  
  all $x \in \mathbb{R}^{d}$, and hence the Carleman property
  \eqref{carledef} for $\overline{k_{t}}$. We defer the proof of the
  continuity of $\overline{k_{t}}$ 
  to the end, but exploit its consequences right now. Jensen's inequality,
  Fubini's theorem and the almost-surely applicable Markov property
  \eqref{markov} yield the estimate
  \begin{eqnarray}
    \| \overline{k_{t}}(x,\cdot) - \overline{k_{t}}(z,\cdot) \|^2_{2}
    & \le & \int_{\mathbb{R}^{d}}\!\d y\; \mathbb{E}\bigl[ 
    |k_{t}(x,y) - k_{t}(z,y)|^{2} \bigr] \nonumber\\
    & = &  \overline{k_{2t}}(x,x) - \overline{k_{2t}}(z,x) - 
    \overline{k_{2t}}(x,z) + \overline{k_{2t}}(z,z)\,,\qquad\qquad
  \end{eqnarray}
  showing that the continuity of $\overline{k_{2t}}$ implies the strong
  continuity of the mapping 
  $\mathbb{R}^{d}\to\mathrm{L}^{2}(\mathbb{R}^{d})$, $x \mapsto 
  \overline{k_{t}} (x,\cdot)$.

  The estimate \eqref{freekernelbound} delivers
  \begin{equation}
    \label{freebound}
    |T_{t}\psi| \le \mathcal{L}_{t} \, \e^{-tH(0,0)} |\psi|
  \end{equation}
  for all $\psi\in\mathrm{L}^{2}(\mathbb{R}^{d})$, where $T_{t}$ is
  defined as in \eqref{uteq}. Consequently,  $T_{t}$ is a bounded
  Carleman operator on $\mathrm{L}^{2}(\mathbb{R}^{d})$. Moreover,
  $T_{t}$  is self-adjoint because of the Hermiticity of
  $\overline{k_{t}}$ and an interchange of integrations thanks to
  \eqref{freekernelbound} and Fubini's theorem. The continuity of 
  any image $T_{t}\psi$ follows from the strong continuity of $\overline{k_{t}}
  (x,\cdot)$ by proceeding along the lines of Eq.\ \eqref{contest} in the
  proof of Lemma~\ref{kate}. 

  Now let $\psi\in \mathrm{L}_{\mathrm{G}}^{2}(\mathbb{R}^{d})$ 
  so that the equality $T_{t}\psi = \mathbb{E} \bigl[ \e^{-tH(A,V)}\psi
  \bigr]$ follows from \eqref{fkieq} and an
  interchange of integrations. This interchange is again 
  allowed by Fubini's theorem and \eqref{freekernelbound}. The
  inequalities \eqref{freekernelbound} and \eqref{elineq} imply that
  $T_{t}\psi \in  
  \mathrm{L}_{\mathrm{G}}^{\infty}(\mathbb{R}^{d})$ for all $\psi\in
  \mathrm{L}_{\mathrm{G}}^{2}(\mathbb{R}^{d})$. Remark~\ref{lpwell} applies
  accordingly. 

  Next we establish the positivity of $T_{t}$. Given any $\psi\in
  \mathrm{L}_{\mathrm{G}}^{2}(\mathbb{R}^{d})$, one deduces from the
  just-proven equality \eqref{Uinterpr}, the estimate
  \eqref{freekernelbound} and Fubini's theorem that $\langle\psi,
  T_{t} \psi\rangle = \mathbb{E} \bigl[\langle\psi, \e^{-tH(A,V)}\psi\rangle
  \bigr]\geq 0$, where the lower bound follows from the positivity of
  $\e^{-tH(A,V^{(\omega)})}$ for $\mathbb{P}$-almost all
  $\omega\in\Omega$. Now, the denseness of
  $\mathrm{L}_{\mathrm{G}}^{2}(\mathbb{R}^{d})$ in
  $\mathrm{L}^{2}(\mathbb{R}^{d})$, the boundedness of $T_{t}$ and the
  continuity of the scalar product yield $\langle\psi,
  T_{t} \psi\rangle \geq 0$ for all $\psi\in
  \mathrm{L}^{2}(\mathbb{R}^{d})$. 
  
  Finally, we turn to the postponed proof of the continuity of the
  mapping $\mathbb{R}^{d}\times\mathbb{R}^{d} \rightarrow
  \mathbb{C}\,, (x,y) \mapsto 
  \overline{k_{t}}(x,y)$. This continuity will follow from
  Lemma~\ref{kmeas} and the dominated-convergence theorem, provided we show 
  \begin{equation}
    \label{stetlemma}
    \mathbb{E} \biggl[ \sup_{x,y\in\mathcal{K}}
    |k_{t}(x,y)|\biggr] < \infty 
  \end{equation}
  for any bounded set
  $\mathcal{K}\subset\mathbb{R}^{d}\times\mathbb{R}^{d}$. In order to
  do so, let us fix $\omega\in\Omega_{0}$ and 
  $x,y\in \mathcal{K}$ arbitrary. By using
  \eqref{kern}, the triangle inequality, Jensen's inequality
  and Fubini's theorem, we get
  \begin{eqnarray}
    |k_{t}^{(\omega)}(x,y)| &\le& (2\pi t)^{-d/2} \int_{0}^{t}
    \frac{\d s}{t} 
    \int\mu_{x,y}^{0,t}(\d b)\; \e^{-t V^{(\omega)}(b(s))} \nonumber\\
    & = & (2\pi t)^{-d/2} \int_{0}^{1}\!\d\sigma \int_{\mathbb{R}^{d}} 
    \!\d z\; g_{\sigma} \bigl(z- m_{x,y}(\sigma)\bigr)  \,
    \e^{- t V^{(\omega)}(z)}\,, \qquad\qquad
    \label{continit}
  \end{eqnarray}
  where the equality follows from an explicit computation with
  $m_{x,y}(\sigma) := x+ (y-x)\,\sigma$ and 
  \begin{equation}
    g_{\sigma}(z) := \frac{\exp\{- |z|^{2} /[2(1-\sigma)\sigma t]\}}%
    {[2\pi(1-\sigma)\sigma t]^{d/2}}\,.  
  \end{equation}
  Next we apply H\"older's inequality with the conjugated exponents $p
  \in ]1,\infty[$ and $p':=(1-p^{-1})^{-1}$ to the integral with
  respect to $z$
  in \eqref{continit}, which yields the upper bound
  \begin{equation}
    \label{hoelderresult}
    \left( \int_{\mathbb{R}^{d}}\!\d z\;
      \e^{- p t V^{(\omega)}(z)} \, \e^{-p|z|} \right)^{1/p}
    \left( \int_{\mathbb{R}^{d}}\!\d z\; \e^{p'|z|}
      \bigl|g_{\sigma}\bigl(z-m_{x,y}(\sigma)\bigr)\bigr|^{p'}
    \right)^{1/p'}.  \quad
  \end{equation}
  The second integral in \eqref{hoelderresult} is bounded
  from above by 
  \begin{equation}
    \label{hoelderbound}
    \quad \e^{p' \max\{|x|,|y|\}} \int_{\mathbb{R}^{d}}\!\d z\; \e^{p'|z|}
    \, |g_{\sigma}(z)|^{p'}
    \le \e^{p' \max\{|x|,|y|\}} \, [(1-\sigma)\sigma t]^{(1-p')d/2}\,
    I_{p'}\,, \quad
  \end{equation}
  where $I_{p'} := (2\pi)^{-d/2} \int_{\mathbb{R}^{d}} \!\d \zeta \, 
  \e^{-p'(|\zeta|^{2} - |\zeta| \sqrt{t})/2} < \infty$ for any $p'>1$. This
  gives the estimate
  \begin{eqnarray}
    \mathbb{E} \biggl[ \sup_{x,y\in\mathcal{K}}
    |k_{t}(x,y)|\biggr]  &  \le &
    (2\pi t)^{-d/2} I_{p'}^{1/p'} \left(\sup_{z\in\mathcal{K}}
      \e^{|z|}\right) \int_{0}^{1}\!\d\sigma [(1-\sigma)\sigma
    t]^{-d/(2p)} \nonumber \\
    && \hspace*{1cm} \times \,\mathbb{E}\left[ \left(
        \int_{\mathbb{R}^{d}}\!\d z\; 
      \e^{- p t V(z)} \, \e^{-p|z|} \right)^{1/p} \right] \,.
    \label{contfin}
  \end{eqnarray}
  The expectation value on the right-hand side of \eqref{contfin} is
  finite for any $p>1$ by Jensen's inequality, property \ass{L} and
  Fubini's theorem. Therefore \eqref{stetlemma} follows from the
  boundedness of $\mathcal{K}$ and by choosing $ p> \max\{1, d/2\}$.
\end{proof}


%
%
\begin{acknowledgment}
  It is a pleasure to thank Vadim Kostrykin and Simone Warzel for
  helpful discussions and comments on the manuscript.
\end{acknowledgment}

%


%

\end{article}
\end{document}